\documentclass[a4paper,11pt]{article}
\pdfoutput=1 % if your are submitting a pdflatex (i.e. if you have
\usepackage{jheppub} % for details on the use of the package, please
\usepackage[T1]{fontenc} % if needed
\usepackage{relsize}
\usepackage{slashed}
\usepackage{ytableau}
\usepackage{amsthm}
\usepackage{caption}
\usepackage{subcaption}
\usepackage[utf8]{inputenc}
\usepackage{cleveref}
\crefformat{section}{\S#2#1#3} % see manual of cleveref, section 8.2.1
\crefformat{subsection}{\S#2#1#3}
\crefformat{subsubsection}{\S#2#1#3}

\usepackage[force]{feynmp-auto}

\usepackage{braket}
\usepackage{filecontents}

\newcommand{\marrow}[5]{%   
    \fmfcmd{style_def marrow#1
    expr p = drawarrow subpath (1/4, 3/4) of p shifted 3 #2 withpen pencircle scaled 0.4;
    label.#3(btex #4 etex, point 0.5 of p shifted 6 #2);
    enddef;}
    \fmf{marrow#1,tension=0}{#5}}
    
\def \12x12 {$\frac{1}{2}\otimes \frac{1}{2}$ }
\def \0x1 {$0\otimes1$ }
\newcommand{\Rlm}[4]{{}_{#1}R_{#2 #3 #4}}
\newcommand{\Rin}[4]{{}_{#1}R^{\rm in}_{#2 #3 #4}}
\newcommand{\sSlm}[3]{{}_{#1} S_{#2 #3}}

\newcommand{\rs}{r_*}
\newcommand{\Binc}[3]{B^{\rm inc}_{#1 #2 #3}}
\newcommand{\Bref}[3]{B^{\rm ref}_{#1 #2 #3}}
\newcommand{\Btrans}[3]{B^{\rm trans}_{#1 #2 #3}}

\newcommand{\diff}[2]  {\frac{d #1}{d #2}}

\newcommand{\pdiff}[2]  {\frac{\partial #1}{\partial #2}}
\newcommand{\spdiff}[2] {\frac{\partial^2 #1}{\partial #2^2}}

\title{Scattering in Black Hole Backgrounds and Higher-Spin Amplitudes: Part I }
\author[a,b]{Yilber Fabian Bautista,}\emailAdd{ybautistachivata@pitp.ca}
\author[a,c]{Alfredo Guevara,}\emailAdd{aguevaragonzalez@fas.harvard.edu}
\author[d]{Chris Kavanagh}\emailAdd{chris.kavanagh@aei.mpg.de }
\author[d]{and Justin Vines}\emailAdd{justin.vines@aei.mpg.de }

\affiliation[a]{Perimeter Institute for Theoretical Physics, Waterloo, ON N2L 2Y5, Canada}
\affiliation[b]{Department of Physics   and  Astronomy, York University, Toronto, Ontario, M3J 1P3, Canada}
\affiliation[c]{Society of Fellows, Harvard University, Cambridge, MA 02138, USA}
\affiliation[d]{Max Planck Institute for Gravitational Physics (Albert Einstein Institute), Am  M\"{u}hlenberg 1, Potsdam 14476, Germany}

\abstract{The scattering of massless waves of helicity $|h|=0,\frac{1}{2},1$ in Schwarzschild and Kerr backgrounds is revisited in the long-wavelength regime. Using a novel description of such backgrounds in terms of gravitating massive particles, we compute classical wave scattering in terms of $2\to 2$ QFT amplitudes in flat space, to all orders in spin. The results are Newman-Penrose amplitudes which are in direct correspondence with solutions of the Regge-Wheeler/Teukolsky equation. By introducing a precise prescription for the point-particle limit, in Part I of this work we show how both agree for $h=0$ at finite values of the scattering angle and arbitrary spin orientation. 

Associated classical observables such as the scattering cross sections, wave polarizations and time delay are studied at all orders in spin. The effect of the spin of the  black hole  on the polarization and helicity of the waves is found in agreement with previous analysis at linear order in spin. In the particular limit of small scattering angle, we argue that wave scattering admits a universal, point-particle description determined by the eikonal approximation. We show how our results recover the scattering eikonal phase with spin up to second post-Minkowskian order, and match it to the effective action of null geodesics in a Kerr background. Using this correspondence we derive classical observables such as polar and equatorial scattering angles. 

This study serves as a preceding analysis to Part II, where the Gravitational Wave ($h=2$) case will be studied in detail.

}

\begin{document} 
\maketitle
\flushbottom

\section{Introduction}
Since the early days of General Relativity, the stability of space-times as dictated by Einstein's field equations has been an important open problem with far-reaching consequences from mathematics to phenomenology. In their seminal work \cite{Regge:1957td} (see also \cite{Chandrasekhar:579245}) Regge and Wheeler (RW) proved that the Schwarzschild black hole was stable under small perturbations by introducing the differential equation 

\begin{equation}\label{eq:RWeq}
    \frac{d^2 R_\ell(r)}{dr_{*}^2} + (E^2 - V_\ell(r))R_\ell(r)=0 \,, \quad V_\ell(r)=\left (1-\frac{2GM}{r}\right)\left[\frac{\ell(\ell+1)}{r^2}+ \frac{2GM(1-h^2)}{r^3}\right]\,,
\end{equation}
governing the radial dependence of a perturbation with spin $h$ and orbital angular momentum $\ell\geq |h|$. Here  $r_{*}$ is the usual tortoise radial coordinate. The Regge-Wheeler equation sets on the same footing the cases of a scalar ($h=0$), neutrino ($|h|=1/2$), electromagnetic ($|h|=1$) and gravitational ($|h|=2$) waves. In a more modern language, the full perturbation is easily encoded in a Newman-Penrose scalar for spin $h$\footnote{In the more standard notation of e.g. \cite{Newman:1961qr}, this corresponds to the top scalar $\boldsymbol{\Psi}_{(-2h)}$ for a helicity-$h$ field.}

\begin{equation}\label{eq:npsc}
    \boldsymbol{\Psi}_h(t,r,\theta,\phi)\propto\frac{1}{r} e^{-i E t}\sum_{l=0}^\infty D_r^h R_{\ell }(r)\,\, {}_{h}Y_{\ell 0}(\theta,\phi)\,,
\end{equation}
 where ${}_{h}Y_{\ell m}$ are spherical harmonics of spin-weight/helicity $h$ \cite{Goldberg:1966uu} and $D_r^h$ is a differential operator trivial for $h=0$. Although closed-form solutions to \eqref{eq:RWeq} are rare, numerical and analytical methods can be implemented to gain insight into rich black hole physics, see e.g. \cite{Berti:2004md,Fiziev:2005ki}. In particular, for scattering situations one is interested in the radiative content of $\boldsymbol{\Psi}_h$, which is encoded in the asymptotic behaviour of $R_{\ell }(r)$  as $r\to \infty$ \cite{Newman:1968uj,futterman88}. In fact, one can gain insight by thinking of $\boldsymbol{\Psi}_h$ as a certain effective wavefunction and identifying \eqref{eq:RWeq} as its associated time-independent Schrödinger equation. This hints that Quantum Mechanics (QM) can be used effectively to describe the perturbations $\boldsymbol{\Psi}_h$ provided a certain classical prescription is implemented.
 
 On the other hand, it is well known that scattering solutions in Quantum Mechanics can be easily recovered by a more modern, second-quantized, scattering amplitude in Quantum Field Theory (QFT). This suggests that scattering governed by \eqref{eq:RWeq} can be described by a QFT amplitude featuring massless states and no-particle production. In \cite{PhysRevD.16.237} this intuition was confirmed and it was shown that such S-matrix is reproduced in the Born approximation by the scattering of scalars ($h=0$), photons ($h=\pm 1$) and gravitons ($h=\pm 2$) off a background current associated to the Schwarzschild metric.

What is the difference between the amplitude computed from the Schrodinger-like equation \eqref{eq:RWeq} and from the QFT picture? For a given order in $GM/r$, corrections to  \eqref{eq:RWeq} can only appear through $\frac{E}{M}\sim \frac{1}{r M}$, where $r$ is conjugate to the momenta. Thus, we can disregard these corrections in the QFT amplitude by implementing the  `classical limit' $E\to \hbar \omega$, $r\to r/\hbar$ and taking $\hbar\to 0$. This is equivalent to a multiple soft limit where the momenta for both real and virtual massless particles scales as $k_i = \hbar \hat{k}_i \to 0$. A multiple soft limit approach has been used in diverse contexts to obtain classical observables, see e.g. \cite{Strominger:2017zoo,Guevara:2018wpp,Guadagnini:2008ha,Laddha:2018rle}, where the classical radiative field usually emerges as a coherent state made of on-shell massless particles.

A purely second-quantized approach would also demand the source in \eqref{eq:RWeq}, i.e. the Schwarzschild black hole, to be represented by an interacting particle in a scattering amplitude. This resonates with a remarkable fact, already suggested in the seminal analysis of \cite{Duff:1973zz}, that classical dynamics of black holes can be obtained by representing them as fundamental massive particles. Such correspondence has recently (re)emerged to gain deep insight into the binary black hole problem \cite{Damour_2016,Cheung:2018wkq, Cheung:2020gyp,Bern:2019nnu,Bern:2019crd,Bern:2021dqo,Bjerrum-Bohr:2018xdl,Cristofoli:2019neg,  Bjerrum-Bohr:2019kec, Bjerrum-Bohr:2021din,Bjerrum-Bohr:2021vuf, DiVecchia:2020ymx,DiVecchia:2021ndb,DiVecchia:2021bdo,Kosower:2018adc, Mogull:2020sak,Kalin:2019rwq,Kalin:2019inp,Levi:2020kvb,Levi:2020uwu,Kalin:2020mvi, Kalin:2020fhe, Damour:2020tta,Damour:2019lcq, Maybee:2019jus,  Arkani-Hamed:2019ymq,Bern:2020buy,Chung:2019duq,Chung:2020rrz, Cachazo:2017jef,Guevara:2017csg,Guevara:2018wpp,Guevara:2019fsj,Aoude:2020onz,Goldberger:2020fot, Goldberger:2017vcg, Goldberger:2017ogt,Bautista:2019tdr, Bautista:2019evw,Herrmann:2021lqe, Herrmann:2021tct,Jakobsen:2021smu, Bern:2020uwk,Jakobsen:2021lvp,Mougiakakos:2021ckm,Cristofoli:2020uzm,Vines:2018gqi,Blumlein:2020znm}. In particular, it has been shown that the perturbative Schwarzschild metric can be directly mapped to a three-point amplitude $A_3$ for a massive particle emitting a graviton \cite{Donoghue:1994dn,Bjerrum-Bohr:2018xdl,Jakobsen:2020ksu,PhysRevD.68.084005,Mougiakakos:2020laz}, where the latter encodes the gauge invariant on-shell modes of the metric. The situation is even more interesting in the case of spinning black holes, where it has been found that the linearized Kerr metric can be mapped to $A_3$ for a massive particle of infinite quantum spin \cite{Guevara:2018wpp,Chung:2018kqs,Arkani-Hamed:2019ymq,Chung:2019yfs}. The fact that the higher-spin particle effectively models the Kerr black hole in a classical regime can be phrased as the property of being \textit{minimally coupled} to gravity, in the sense that it has a smooth high-energy limit $M\to 0$ \cite{Arkani-Hamed:2017jhn}.

Higher multiplicity amplitudes, associated to a massive particle emitting $n>1$ gravitons or massless particles, have also recently gathered attention in connection to higher perturbative orders of the two-black hole problem. More precisely, four-point amplitudes emerge naturally in unitarity cuts when evaluating both conservative \cite{Guevara:2018wpp,Guevara:2019fsj,Bern:2019crd,Bern:2019nnu,Bern:2021dqo,Bern:2020buy,Bjerrum-Bohr:2013bxa,Cachazo:2017jef} and radiation \cite{ Bautista:2019tdr,Herrmann:2021lqe,Herrmann:2021tct}   two-body effects.  In that context, a concern was raised in  \cite{Arkani-Hamed:2017jhn,Guevara:2018wpp,Chung:2018kqs,Chung:2019duq,Aoude:2020onz}  regarding the appropriate higher-spin amplitude $A_4$ that could model the interaction of spinning black holes. In fact, at four points the aforementioned notion of minimal coupling  breaks down, as any description of higher-spin particles contains $1/M$ singularities.\footnote{In other words such amplitudes hold only at low energies energies $E\ll M$. This has been observed to be intimately tied to causality violation in classical regimes \cite{PhysRevD.46.3529, Deser:2001dt,Deser:2000dz,Cucchieri:1994tx}.} Because of this, no obvious candidate arises to encode the Kerr black hole in a four-point amplitude.

In this two-part work, we shed light on the the four-point amplitudes associated to Schwarzschild and Kerr black holes. For this, we first define a suitable classical limit of the four-point QFT amplitudes appropriate for the wave scattering process. We then establish a map from such classical amplitude to the radiative NP scalar appearing in \eqref{eq:RWeq}.  Furthermore, to tackle the case of spinning black holes/particles we appeal to the generalization of the Regge-Wheeler equation for the Kerr black hole, introduced long ago by Teukolsky \cite{1973ApJ...185..635T,articleteu2,1974ApJ...193..443T}. This equation expresses a remarkable simplification in the treatment of Kerr perturbations in contrast with generic backgrounds, and stands as a cornerstone of the modern black hole perturbation theory (BHPT) framework (see e.g. \cite{Pound:2021qin} for a recent review). In this paper, Part I, we will consider the scattering of waves of helicity $h=0,1/2,1$ and extend previous BHPT results \cite{PhysRevD.16.237,Westervelt:1971pm,Doran:2001ag,Dolan:2007ut,Matzner1968,Chrzanowski:1976jb, PhysRevD.13.775} to all orders in spin, by introducing novel Feynman diagrammatic rules for our QFT higher-spin amplitudes. The matching sheds new light on different observable effects that are well-known for scattering in Kerr: For instance, we derive how the spin of the black hole induces polarization on the waves and we provide a new argument for classical conservation of their helicity. Furthermore, unlike the $h=2$ case which will be studied in Part II, these amplitudes do not contain $1/M$ singularities and are in a sense directly connected to the minimal-coupling three-point vertices, thereby naturally extending such a notion to four-points.

To match the NP scalars to amplitudes at weak coupling we will consider the long-wavelength regime $GM\omega = \pi r_S/\lambda \ll 1$ in which the black hole behaves as a point-like object. In the spinning case, we find that a natural prescription in such limit is to consider the leading order in $GM/a\to 0$ of solutions to the Teukolsky equation, which introduces an analytic continuation of the physical region $GM/a\geq 1$ . The result then yields a series of multipole corrections in powers of $a\omega=2\pi a/\lambda$, reflecting that the scattered wave can indeed probe the internal structure of the spinning body.

As an important consistency check we will consider the geometrical optics, i.e. eikonal, approximation \cite{tHooft:1987vrq,Amati:1987wq,Amati:1987uf,Amati:1990xe,Kabat:1992tb,Verlinde:1991iu,Camanho:2014apa,Akhoury:2013yua,Melville:2013qca,DiVecchia:2019kta,brittin1959lectures}, arising at small scattering angle and large orbital angular momentum in our classical amplitudes. It can already be argued from the potential in \eqref{eq:RWeq} that the $\ell\to \infty$ regime of the  wave scattering process is universal (independent of $h$) and is controlled by the effective potential of a null geodesic. Indeed, we will show that by exponentiating the QFT amplitudes for $h\leq 2$ in the eikonal limit we can recover the scattering observables of a null geodesic, such as Shapiro time delay, establishing a wave-particle duality on the classical side. This provides a three-way crosscheck: On the one hand we match the results from the Teukolsky equation to the eikonal phase from  spinning QFT amplitudes, up to 1-loop. On the other hand, from the eikonal phase we derive two different scattering angles, namely polar and equatorial, that then match those computed from null geodesics of a Kerr background.

This paper is organized as follows. In \cref{sec:2} we consider the scattering of waves of $h=0,1/2,1,2$ in the Schwarzschild background. We define the classical limit of 4-point amplitudes and establish the link with the Newman-Penrose scalar. In \cref{sec:Section3} we then extend the construction to the Kerr background/higher-spin amplitudes. We also revisit observable classical effects such as spin-induced polarization of the wave and helicity violation/conservation. In \cref{teukolsky} we introduce the BHPT framework and present the results for the infinite series of partial waves in the aforementioned point-like limit. After projecting the previous NP scalars into such partial waves we obtain a perfect match. Finally, \cref{Eikonal-app} introduces the particle-wave duality in the eikonal regime and presents the matching to null geodesics. In Appendix \ref{Teuk-App} we revisit the solutions of the Teukolsky equation using BHPT, whereas Appendix \ref{genhel} provides a derivation of eikonal universality from 4-pt amplitudes.

\section{Scattering of Waves in Schwarzschild Backgrounds}\label{sec:2}

From the perspective of scattering amplitudes in QFT, the Schwarzschild Black Hole (SBH) of $r_S=2GM$ can be modeled by a scalar particle of mass $M$ which is minimally coupled to gravitons \cite{Duff:1973zz}. More generally, it has been observed the scalar, neutrino, electromagnetic and gravitational wave perturbations in such background can be modeled by the classical limit of massless particles of helicity $h\leq2$ \cite{Guadagnini:2008ha,PhysRevD.16.237}.

\begin{figure}
\begin {center}
\includegraphics[width=7truecm]{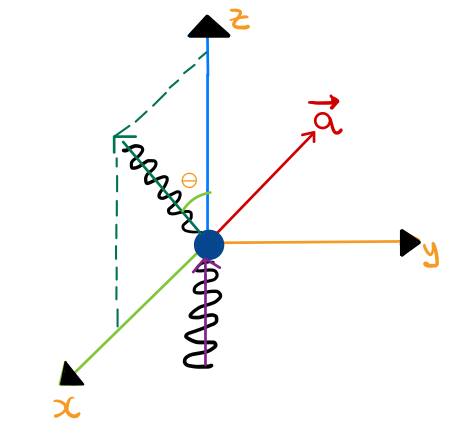} 
\end{center}
\caption{ Wave scattering process in an Amplitude setup. An incoming plane wave  traveling along the $z$-axis hits a  BH at  rest; the wave gets scattered with outgoing  momentum lying in the $x-z$ plane.  We have introduced a spin vector $\vec{a}$, oriented in a generic direction, in  preparation for the process of   wave scattering off Kerr BH,  treated in \cref{sec:Section3}. 
}
\label{fig:ampl}
\end{figure}

In order to define the classical piece of the QFT amplitude and link it to wave scattering we proceed as follows. The null momenta of the massless particles, $k_i$, is to be identified with the classical wavenumbers, $\hat{k}_i$, corresponding to the direction of wave propagation. Thus, this leads us to write

\begin{equation}\label{hscaling}
    k_i = \hbar \hat{k}_i  \,, [\hat{k_i}]=[1/L]\,,
\end{equation}
as $\hbar\to 0$. This scaling will be sufficient for the QFT scattering amplitudes we are interested in, involving a single matter line (see also \cite{Kosower:2018adc,Bautista:2019evw,Bautista:2019tdr}). For such case this also implies that internal massless momentum $q=\sum_i k_i$ has the same behaviour, $q=\hbar \hat{q}$. This is precisely the $\hbar$ deformation explained in the introduction: The scaling $E \to \hbar \omega$ corresponds to the frequency of external momentum whereas $r\to \hbar^{-1} r$ corresponds to the scaling (massless) momentum transfer $q\sim 1/r$.

Consider now the object $A_4^{h}$ representing the $D=4$ four-point scattering amplitude\footnote{The scattering amplitude is defined in such a way that the momentum
conservation delta function is striped away, and overall factors of $i$ are removed. We work in the mostly minus signature for the metric, $\eta = \rm{diag}(1,-1,-1,-1)$.} of two massive scalar legs of momenta $p_{1}$ and $p_{4}$ and two massless legs
of momenta $k_{2}$ and $k_{3}$ (with certain helicity $\pm h$). In the classical interpretation, the massive momenta will be associated to initial and final states of the black hole, whereas the massless momenta represent the incident and scattered wave. The classical limit is the leading order term  in the $\hbar\rightarrow 0$ expansion
of the amplitude,

\begin{equation}
\langle A_{4}\rangle :=\lim_{\hbar\rightarrow0}A_{4}\,.\label{eq:classical amplitude}
\end{equation}
To make contact with BHPT we choose to evaluate it in the reference frame for which the black hole (BH) is
initially at rest and the scattering process is restricted to the $x-z$ plane. By adopting the scaling given in \eqref{hscaling} and the rest frame for $p_1$, the momenta of the particles read (see  Figure \ref{fig:ampl})
\begin{equation}
\label{eq:kinematics}
    \begin{split}
    p_{1}^{\mu} & =(M,0,0,0),\\
k_{2}^{\mu} & =\hbar \omega(1,0,0,1),\\
k_{3}^{\mu} & =\frac{\hbar \omega(1,\sin\theta,0,\cos\theta)}{1+2\frac{\hbar \omega}{M}\sin^{2}(\theta/2)},\\
p_{4}^{\mu} & =p_{1}^{\mu}+k_{2}^{\mu}-k_{3}^{\mu},  
    \end{split}
\end{equation}
 where the form of the energy of the outgoing wave $k_3$ is fixed by the on-shell condition for the outgoing massive particle. The independent kinematic invariants are
\begin{equation}\label{eq:sandtclassical}
    \begin{split}
     s & =(p_1+k_2)^2=M^{2}\left(1+2\frac{\hbar \omega}{M}\right)\,,\\
t & =(k_3-k_2)^2=-\frac{4\hbar^2 \omega^{2}\sin^{2}\left(\theta/2\right)}{1+2\frac{\hbar \omega}{M}\sin^{2}(\theta/2)}\,.   
    \end{split}
\end{equation}
It will also be convenient to introduce the optical parameter:

\begin{equation}\label{eq:xidef}
    \xi^{-1}:=-\frac{M^2 t}{(s-M^2)(u-M^2)}= \sin^2(\theta/2) \,.
\end{equation}

We can now distinguish the following expansions arising in the scattering of waves off a spinless massive object:

\begin{itemize}
    \item The classical expansion in the dimensionless combination $\epsilon=2 GM\omega$. This is in correspondence to the perturbative (loop) expansion in the QFT amplitude and classically corresponds to a weak-field expansion. We will work at leading order in $GM\omega$, which corresponds to tree-level amplitudes. As anticipated, the frequency $\omega$ is also the energy of the momentum transfer $k_3-k_2\sim 1/r$: This means that the classical combination is $GM\omega \sim GM/r$ and hence BH-horizon effects should not be relevant in the weak-coupling limit. 
    \item Quantum corrections suppressed by $\frac{\hbar \omega}{M}$. These corrections appear in $A_4^h$ as powers of $\frac{(s-M^2)}{M^2}$ or $\frac{t}{M^2}$ and will be dropped for our matching. We see that the classical deformation $\frac{\hbar \omega}{M}\to 0$ can be understood as a limit in which massless particles become soft or equivalently the massive particle becomes very heavy. Note that the opposite (high-energy/massless) limit $\frac{\hbar \omega}{M}\to \infty$ for which our effective description can break down as explained in the introduction, simply lies outside the classical regime.
    \item The eikonal (geometrical optics) expansion as $\xi^{-1}\sim \theta \to 0$, which will be studied in detail in Section \ref{Eikonal-app}. The eikonal has been recently exploited to extract classical observables of the two-body problem via a WKB-type analysis  \cite{Bjerrum-Bohr:2016hpa,AccettulliHuber:2020oou,Cristofoli:2020hnk,Guevara:2018wpp} and we will obtain related results that link wave scattering to unbounded null geodesics.  In this regime $t \ll s-M^2 \ll M^2$ and hence only the t-channel factorization $A^h_4\sim \frac{1}{t} A_3 A'^{h}_3$ contributes to the amplitude. Nevertheless, note that for generic wave scattering we are able to extract the classical piece at all orders in $\xi^{-1}$ and not only the eikonal, which demands to include all channels and contact terms in the evaluation of \eqref{eq:classical amplitude}. This non-linear dependence in $\theta$ is crucial to match the partial amplitudes $A_{\ell m}Y_{\ell m}(\theta,\phi)$ resulting from the BHPT formalism. This, therefore, shows the massless wave scattering off Kerr carries more information that the massless limit of a  massive particle  scattering off Kerr, which has built in the assumption of small scattering angles from the beginning. We expect therefore the  two scenarios only overlap in the eikonal regime. 
    
\end{itemize}
Further considerations are required for the spin expansion, see Section \ref{sec:Section3}. In a nutshell, we shall adopt the classical limit simply by setting $\hbar = 1$ and then taking the "multisoft" limit $\omega \to 0$ at a fixed power of $G$. From a classical perspective, this corresponds to the long-wavelength regime $2GM\omega =\pi r_S/\lambda \ll 1\,$ \cite{1964PhDT........64H, Matzner1968,    Mashhoon1973zz,PhysRevD.10.1059}, for which the internal structure of the classical object (including e.g. dispersive effects/quasinormal modes) cannot be probed at leading order. Such information would in principle require corrections in $GM\omega$ corresponding to loop diagrams. Nevertheless, for the case of spinning bodies we will see that there are indeed tree-level corrections $(a\omega)\sim 1/\lambda$ that effectively probe the internal structure.

The classical amplitude as defined in \eqref{eq:classical amplitude} transforms covariantly under a $U(1)$ phase, corresponding to the little group of the massless $|h|\leq 2$ states. We will take care of the phase ambiguity when performing the matching to classical scattering. As suggested by related results in quantum and classical scattering \cite{Bjerrum-Bohr:2016hpa,1964PhDT........64H, Matzner1968, Mashhoon1973zz,PhysRevD.10.1059} the matching may also be performed unambiguously in terms of the unpolarized  classical differential
cross section. For that
we use the well known formula for the differential cross section
in QFT, and then proceed to take its classical limit according to our prescription. Let us assume  that the incoming massless particles have fixed helicity $h>0$, whereas the outgoing massless particle can in general have a different helicity $h^\prime = \pm h$.
Then, the unpolarized differential cross section will be given by 
\begin{equation}\label{eq:crsec}
d\sigma=\sum_{h^\prime}\frac{|A_{4}(h\rightarrow h^\prime)|^{2}d{\rm LIPS}_{2}}{2E_{1}2E_{2}|\vec{v}_{1}-\vec{v}_{2}|}\,,
\end{equation}
where the sum runs over all the polarization states for the outgoing  massless particle, and the two particle Lorentz invariant phase space has the simple
form
\begin{equation}
    d{\rm LIPS}_{2}=\frac{s-M^{2}}{32\pi^{2}s}d\Omega\,.
\end{equation}
Noting that
in the classical limit $k_{3}^{\mu}\rightarrow  \omega(1,\sin\theta,0,\cos\theta)$, the differential cross section simply
becomes
\begin{equation}
\frac{d\langle\sigma\rangle}{d\Omega}=\sum_{h^\prime} \frac{|\langle A_{4}(h\rightarrow  h^\prime)\rangle|^{2}}{64\pi^{2}M^{2}}\,.\label{eq:classical cross section}
\end{equation}
The  impinging wave can also be unpolarized. In such case, the helicity states for both in and out waves allow us to define the elements of the scattering matrix as follows 
\begin{equation}\label{eq:scattering-matrix-helicities}
A_{4}^{h}=\left[\begin{array}{cc}
A_{++}^{h} & A_{+-}^{h}\\
A_{-+}^{h} & A_{--}^{h}
\end{array}\right],
\end{equation}
where the sub-indices denote the polarization  of the incoming and outgoing wave respectively, and $h$ denotes the nature of the wave\footnote{ It is understood that for $h=0$, the scattering  matrix is one dimensional.}. We associate $+h$ ($-h$) states with circular left (right) wave polarizations. Motivated by the discussion of wave polarization in the next sections, we will refer to the diagonal elements $A^h_{++},A^h_{--}$ as \textit{helicity preserving} amplitudes, and to the off diagonals $A^h_{+-},A^h_{-+}$ as \textit{helicity reversing}. An important caveat here is that the helicity of particle $k_3$ appears flipped with respect to somewhat standard conventions: As $k_3$ is outgoing with helicity $h'$ it is equivalent to an incoming particle with helicity $-h'$. Writing the scattering matrix in this fashion will become useful when we discuss the effect of the BH on the polarization of the impinging wave, as we will do in the next sections.

\subsection{Scalar Waves}

To illustrate how the computation of the amplitude works, let us proceed
to reproduce the result for the classical scattering of a massless
scalar wave off the SBH. To compute the scattering amplitudes throughout this section we will use the Feynman rules derived in the harmonic gauge, see e.g. \cite{Holstein:2008sx}. The metric perturbation is defined as
\begin{equation}\label{eq:pertmetric}
    g_{\mu \nu}=\eta_{\mu\nu}+\kappa h_{\mu\nu}\,,
\end{equation}
with $\kappa^2=32\pi G$. The gauge condition is

\begin{equation}\label{eq:harmgauge}
    \partial^{\nu} h_{\mu\nu}-\frac{1}{2}\partial_\mu h^{\nu}_{\,\,\nu}=0\,.
\end{equation}

To leading order in $G$ there is only one Feynman diagram
contributing to the amplitude for the scattering the wave off the SBH. This diagram  corresponds to the one graviton exchange
between one massive and one massless scalar. To compute this amplitude we use the  minimal coupling
3-vertex for two scalars legs and one graviton  
\begin{equation}
V_{\phi\phi h}^{\mu\nu}=-i\kappa\left[\mathcal{P}_{\mu\nu\alpha\beta}p_{1}^{\alpha}p_{4}^{\beta}+\frac{M}{2}\eta^{\mu\nu}\right],\label{eq:vertex scalar scalar graviton}
\end{equation}
where $\mathcal{P}_{\mu\nu\alpha\beta}$ is  the trace-reverser operator\footnote{It satisfies $\mathcal{P}_{\mu\nu\alpha\beta}\mathcal{P}^{\alpha\beta\rho\sigma}=\delta_{\mu}^{(\rho}\delta_{\nu}^{\sigma)}$.}
\begin{equation}
  \mathcal{P}_{\mu\nu\alpha\beta}=\frac{1}{2}[\eta_{\mu\alpha}\eta_{\nu\beta}+\eta_{\mu\beta}\eta_{\nu\alpha}-\eta_{\mu\nu}\eta_{\alpha\beta}]\,.  
\end{equation}
The vertex \eqref{eq:vertex scalar scalar graviton} can be also used for the massless scalar by setting $M=0$. We will also need the graviton propagator,
\begin{equation}
D_{\mu\nu\alpha\beta}=\frac{i\mathcal{P}_{\mu\nu\alpha\beta}}{q^{2}+i\epsilon}\,.\label{eq:propagator graviton}
\end{equation}
When working with tree-level amplitudes we can ignore the contour $i\epsilon$ prescription. The computation of the tree-level scattering amplitude is straightforward
and leads to
\begin{equation}
 A_{4}^{h=0}=\frac{i\kappa^{2}p_{1}{\cdot} k_{2}}{2q^{2}}\left[2p_{1}{\cdot}k_{2}+q^{2}\right]\,,
\end{equation}
where $q=k_{3}-k_{2}$ is the momentum transfer. Using the kinematics $(\ref{eq:kinematics})$, the classical
amplitude $(\ref{eq:classical amplitude})$ takes the simple form

\begin{equation}\label{eq:scalara40}
\langle A_{4}^{h=0}\rangle=\frac{i\kappa^{2}(p_{1}{\cdot} k_{2})^2}{q^{2}}=\frac{\kappa^{2}M^{2}}{4\sin^{2}\left(\theta/2\right)},
\end{equation}
In the next section we will show that the Newman-Penrose amplitude, obtained from the scalar \eqref{eq:npsc} as $r \to \infty$, indeed takes the form

\begin{equation}
    \boldsymbol{\Psi}_0\sim \frac{1}{r} \langle A_{4}^{h=0}\rangle e^{-i\omega (t-r)}\,.
\end{equation}
This definition then will naturally extend to waves carrying helicity.

By virtue of equation $(\ref{eq:classical cross section})$
we can directly obtain the classical differential cross section for
the scattering of a scalar wave off the SBH:

\begin{equation}\label{eq:crsscalar}
    \frac{d\langle \sigma \rangle}{d\Omega}= \frac{G^2 M^2}{\sin^{4}\left(\theta/2\right)}\,.
\end{equation}
This is nothing but the gravitational version of the standard Rutherford cross-section. It agrees with the seminal derivation given in \cite{Matzner1968,Chrzanowski:1976jb, PhysRevD.13.775}, see also \cref{sec:toyexample}. The divergence $\approx \frac{16  G^2 M^2}{\theta^4}$ for forward scattering $\theta\to0$ is a universal feature in all the  examples we will study. It follows from the long-range nature of gravity. Indeed, it is well known that Newtonian interaction yields the classical scattering angle $\theta=4G M/b$ as a function of the impact parameter $b$. This means that the scattering cross-section should behave as

\begin{align}\label{eq:newtonian}
    d\langle \sigma \rangle &= b db d\phi \,,\nonumber \\
    &\approx \frac{16 G^2 M^2}{\theta^3}d\theta d\phi\,, \nonumber \\
     &\approx \frac{16 G^2 M^2}{\theta^4}d\Omega\,.
\end{align}

To close this section, we note that the classical amplitude \eqref{eq:scalara40} for the scattering of scalar
waves can be reproduced solely by its the residue as $t\to 0$, corresponding to a three-point factorization. This in 
general does not hold true for waves of higher helicity $|h|>0$. Roughly, this is because in such case the factor $(s-m^2)^{2}$ can appear in the residue as $t\to 0$, for which the combination $(s-m^2)^{2}/t $  has the same order in $\hbar$ as a contact term $\mathcal{O}(t^0)$. This leads to the important conclusion that contact terms in the momentum transfer $q$, which in position space are localized along the BH singularity at $r=0$, can indeed contribute to wave scattering.\footnote{Dropping contact terms in $t$ raised some doubts about the gauge invariance of the  result of \cite{PhysRevD.16.237} when considering the scattering of gravitational waves off the  SBH. It also  led to discrepancies in the literature results for the induced polarization for the scattering of a gravitational wave (GW) off the Kerr black hole to linear order in the spin \cite{Guadagnini:2008ha,Dolan:2008kf}. We will comment on this issue in \cref{sec:polapp} and in \cite{BCGV}.} 

\subsection{Relation to Scattering in Schwarzschild}\label{sec:relsch}

Before we study the case of waves with helicity it is instructive to demonstrate how the above computation relates to a scalar wave scattering off the Schwarzschild black hole.

For a scalar wave the RW equation \eqref{eq:RWeq} given in the introduction is nothing but the Klein-Gordon equation defined on a Schwarzschild background. The Born approximation of the RW equation ($GM\omega \to 0$), which should connect to tree-level amplitudes, can therefore be obtained from the Klein-Gordon equation using the metric of the form \eqref{eq:pertmetric} (which is linearized in $\kappa$). Using the harmonic gauge \eqref{eq:harmgauge} the scalar wave equation becomes

\begin{equation}
    \eta^{\mu \nu}\partial_\mu \partial_\nu  \psi =-\kappa  h^{\mu \nu}  \partial_\mu \partial_\nu  \psi+\mathcal{O}(\kappa^3)\,,
\end{equation}
In contrast with the previous approach here $h_{\mu \nu}$ is not dynamical, but instead the fixed background whose explicit form will not be needed.
Regarding the RHS as an interaction potential, scattering can be studied to leading order in $\kappa$ by implementing the corresponding Lippmann-Schwinger equation. That is, we expand the field into a combination of an incoming plane wave and a scattered wave as follows. Setting $k^2=0,k^0=\omega$, we write

\begin{equation}
\psi(x)= e^{ik\cdot x} \phi^{\textrm{PW}}+e^{-i\omega t} \phi^{\textrm{S}}(x_i) \,,
\end{equation}
so that we obtain the three-dimensional equation

\begin{equation}\label{eq:wveqs}
    (\omega^2 + \vec{\nabla}^2) \phi^{\textrm{S}}(\vec{x})= \kappa  k^{\mu}k^{\nu} h_{\mu \nu}(\vec{x}) e^{i \vec{k}{\cdot} \vec{x}} \phi^{\textrm{PW}} + \mathcal{O}(\kappa^3)\,.
\end{equation}
This is the leading Born approximation. Setting $\phi^{\textrm{PW}}=1$ for simplicity, we define the classical current as given by the Fourier transform of the source in the RHS

\begin{align}
    j(\vec{k}')=&  \int \frac{d^3 \vec{y}}{(2\pi)^3} e^{-i \vec{k}'{\cdot} \vec{y}} \times \left(\kappa  k^{\mu}k^{\nu} h_{\mu \nu}(\vec{y}) e^{i\vec{k} {\cdot}\vec{y}} \right)\nonumber  \,,\\
    =&  \kappa  k^{\mu}k^{\nu} \bar{h}_{\mu \nu}(\vec{q})  \,,\label{eq:scur}
\end{align}
where $\vec{q}=\vec{k}'-\vec{k}$ is the 3-momentum transfer and $\bar{h}_{\mu\nu}(\vec{q})$ is the three-dimensional Fourier transform of the linearized Schwarzschild metric.  It can be read off from equations \eqref{eq:justin position } -\eqref{eq:justin vertex classical} below with $u=(1,0,0,0)$ and $q^0=0$:

\begin{equation}\label{eq:Fouh}
    \bar{h}_{\mu\nu}(q)= \frac{\kappa M^2}{q^2} \mathcal{P}_{\mu\nu\rho\sigma} u^\rho u^\sigma\,.
\end{equation}
It then becomes clear that this can be interpreted as an effective 3-point vertex including the graviton propagator, and that \eqref{eq:scur} is nothing but the t-channel diagram leading to the classical amplitude \eqref{eq:scalara40}

\begin{equation}
    j(k')=\kappa  k^{\mu}k^{\nu} \bar{h}_{\mu \nu}(q)  = \langle A_4^{h=0}\rangle\,,
\end{equation}
with $p_1=Mu$. The condition $q^0=0$ is just the on-shell condition $M^2=(p_1+q)^2 \Rightarrow p_1\cdot q=0$ in the classical limit. It implies that \eqref{eq:Fouh} is indeed in the harmonic gauge $q^\mu \bar{h}_{\mu\nu}=\frac{1}{2} q_\nu \bar h$. Because $\bar{h}_{\mu\nu}$ is off-shell, we note that \eqref{eq:Fouh} has an ambiguity given by contact terms which are regular in $q^2$, as anticipated. More precisely, such terms Fourier transform to $\delta^3(\vec{x})$ terms in position space and therefore are invisible in the Schwarzschild metric. These terms are important and can indeed contribute to the 4-point amplitude, as we will discuss, but for now we assume they are absent.

Having elucidated the relation between the classical current and the amplitude, we proceed to compute the Newman-Penrose field. We define it as the scattered part of the scalar:

\begin{equation}
      \boldsymbol{\Psi}_0:= e^{-i\omega t} \phi^S(\vec{x})\,,
\end{equation}
where $\phi^S(x_i)$ solves the wave equation \eqref{eq:wveqs}. Using the source \eqref{eq:scur}, the solution is given by 

\begin{equation}
    \boldsymbol{\Psi}_0(\vec{x},t)= e^{-i\omega t} \phi^S(x_i)=\int \frac{d^3 \vec{k}}{(2\pi)^3} e^{i(\vec{k}{\cdot} \vec{x}-\omega t)} \frac{j(k)}{\omega^2-\vec{k}^2}\,.
\end{equation}
As expected,  this is just the standard relation between the current and the sourced field (we can assume $j(k)$ to be a function of the 4-vector $(\omega,\vec{k})$). Note that integration in $\omega$ is not performed since we are studying scattering in frequency domain, not in time domain. The radiative data encoded in $\boldsymbol{\Psi}_0(\vec{x},t)$ can be extracted by evaluating it at null infinity via $r=|\vec{x}|\to \infty$. In such limit the integral is customary and easily evaluated using one of its saddle-points (see e.g. Exercise 4 in \cite{Strominger:2017zoo})
we obtain
\begin{eqnarray}\label{eq:psioutin}
\boldsymbol{\Psi}_0^{\textrm{rad}} (\vec{x},t)\sim  \frac{e^{i\omega (r-t) }}{ r}\times j(k)\,.
\end{eqnarray}
where $\vec{k} = \omega \vec{x}/r$ is fixed as pointing to a direction in the celestial sphere. This is the fundamental relation between the NP components and the amplitude: It links the radiative components of the field, given by the NP scalar, to the phase space of massless particles given by the classical amplitude (see \cite{Frolov:1977bp,Ashtekar:1987tt,Strominger:2013lka} for appropriate characterizations of the phase space). Here we have identified $j(k)$ as the classical limit of a quantum current.\footnote{Indeed, this is closely related to the construction of \cite{Strominger:2013lka} and references therein. The identification of quantum and classical currents there is obtained by considering coherent states, see also \cite{CGKO}.} In the next sections we will deal with its natural extension to spinning waves.

Some comments are in order. On the one hand, the argument can be easily seen to make no assumption about the specific form of the metric, which in particular could have been that of a (linearized) spinning body. Thus, our argument relating the scalar amplitude to the NP scalar indeed holds for a Kerr background as well. On the other hand, the Born approximation used here does not take into account several effects that are indeed present in the RW equation: First, it ignores the long-distance drag of the gravitational force, signaled by the presence of $r^*=r+\mathcal{O}(G)$ in \eqref{eq:RWeq}. Indeed, as explained in \cref{eq:asymtpwa} the fall-off of the scattered wave should be $\phi^S\sim e^{i\omega r*} /r$ instead of \eqref{eq:psioutin}. This however does not spoil the relation of the scattering amplitude to the classical current, which is independent of $r$. Second, the Born approximation ignores an overall IR phase known as the Newtonian (or Coulombian) phase. This phase does not contribute to the cross-section, but can be easily reintroduced through the eikonal framework as done in \cref{Eikonal-app}. In the end, our aim is to show that even though these effects are built-in in the RW or Teukolsky equations, we can still match their solutions to our QFT amplitudes if we define a suitable `Born limit' $GM\omega \to  0$ for the former. This will be particularly powerful for waves with spin (such as GW) as the corresponding non-linear wave equations turn out to be considerably more complicated than the Klein-Gordon equation, whereas the RW/Teukolsky system has essentially the same complexity for all helicities.

\subsection{Spinning Waves of $h\leq 1$}
Our next aim is to relate the scattering of spinning waves off the Schwarzschild black hole to the classical limit of the corresponding QFT amplitudes. We will use the electromagnetic case $h=1$ as our guiding example. Since the observable amplitude carries helicity dependence, it is convenient to project it into helicity components.  These are precisely given by the NP scalars. We will also incorporate the limiting procedure of the previous section to present the correspondence between quantum and classical helicity amplitudes.\footnote{A related comprehensive treatment of the classical limit in the context of the Kosower-Maybee-O'Connell formalism is to be given in \cite{CGKO}.}

To begin with, we note that we can decompose the full metric \eqref{eq:pertmetric} into the null tetrad 
\begin{equation}\label{eq:gnullex}
    g_{\mu \nu} =- 2m_{(\mu}\bar{m}_{\nu)} + 2l_{(\mu}n_{\nu)}\,,
\end{equation}
where the only non-vanishing inner products are $  l\cdot n =- m\cdot \bar{m}=1$. In the linearized regime $G\to 0$, such that $g_{\mu\nu}\to \eta_{\mu\nu}$, the null tetrad becomes
\begin{equation}\label{eq:nlltet}
    \begin{split}
    l^{\mu} &\to  (1,\sin \theta \cos \phi,\sin \theta \sin \phi,\cos\theta)\,, \\
      n^{\mu} &\to \textcolor{black}{-}\frac{1}{2}(-1,\sin \theta \cos \phi,\sin \theta \sin \phi,\cos\theta) \,, \\
       m^{\mu} &\to  \frac{1}{\sqrt{2}}(0,\cos \theta \cos \phi - i\sin\phi ,\cos \theta \sin \phi+i\cos\phi,-\sin\theta)\,,   \\
       \bar{m}^{\mu} &\to  \frac{1}{\sqrt{2}}(0,\cos \theta \cos \phi +  i\sin\phi ,\cos \theta \sin \phi-i\cos\phi,-\sin\theta)\,.
    \end{split}
\end{equation}
Note that there is an isometry subgroup $U(1)\subset SO(3,1)$ in $\eqref{eq:gnullex}$ which rotates 
\begin{equation}\label{eq:heltr}
    m^{\mu}\to e^{i\alpha} m^{\mu}, \bar{m}^{\mu}\to e^{-i\alpha} \bar{m}^{\mu}\,,
\end{equation}
This transformation is the starting point of the formalism of Newman and Penrose \cite{Newman_1962}, and characterizes the helicity of the states defined at null infinity. Furthermore, if we identify the flat limit of $l^\mu$ with a normalized massless momentum, $\hat{k}^\mu$, the conjugate pair $m^\mu,\bar{m}^\mu$ may be interpreted as the corresponding polarization vectors $\epsilon^+,\textcolor{black}{-}\epsilon^-$ of helicity $\pm 1$ respectively, satisfying 
\begin{equation}
    \hat{k}\cdot \epsilon^+=\hat{k}\cdot \epsilon^-=0,\,\,\,\,\epsilon^+\cdot\epsilon^- = 1\,.
\end{equation}
Following the argument of the previous section, we now consider the asymptotic EM field for a mode with $k=\omega \hat{k}$\footnote{Here we can ignore the Coulomb dragging which appears in the presence of long-range interactions. We will come back to this point in   \cref{sec:toyexample} and \cref{sec:pwtoeikonal} .}

\begin{equation}\label{eq:classical_A_photons}
   A^{\textrm{rad}}_\mu (x) \propto j_\mu(k) e^{ik\cdot x}\,,
\end{equation}
where as explained only the on-shell modes, i.e. those with $k^2=0$, will yield radiation flux and hence can be associated with the phase space of massless particles. Again, as we approach future null infinity $t+r\to \infty$ only the modes with 3-momentum $\vec{k}$ aligned with $\vec{x}$ will contribute to such flux. That is to say we will take 
\begin{equation}\label{eq:saddle}
\frac{\vec{k}}{\omega}=\frac{\vec{x}}{|\vec{x}|} = (\sin \theta \cos \phi,\sin \theta \sin \phi,\cos\theta)   \Longrightarrow k^\mu \to \omega l^\mu\,,
\end{equation}
in the $G\to 0$ limit, as anticipated. The classical scattering amplitude is described by NP scalar in \eqref{eq:npsc}, which in this case is defined as \cite{futterman88}

\begin{eqnarray}\label{eq:psiphoton}
    \boldsymbol{\Psi}_{-1}^{\textrm{rad}}&:=& -\bar{m}^\mu n^\nu (\partial_\mu A_{\nu}^{\textrm{rad}}- \partial_\nu A_{\mu}^{\textrm{rad}})\,,
\end{eqnarray}
carrying weight $h=-1$ under the helicity transformation \eqref{eq:heltr}, with the conjugate case defined similarly. For gravitational scattering we know that $A_{\mu}^{\textrm{rad}}=\mathcal{O}(G)$, hence hereafter we can consider the null tetrad at the strict flat limit \eqref{eq:nlltet}. Further using (\ref{eq:classical_A_photons}-\ref{eq:saddle}) we have

\begin{eqnarray}\label{eq:psi2rad}    
   \boldsymbol{\Psi}_{-1}^{\textrm{rad}}  &\propto& -\epsilon_{-}^\mu n^\nu (k_\mu j_{\nu}(k) - k_\nu j_{\mu}(k))e^{\omega l\cdot x} = \omega \, \epsilon_-^\mu j_\mu (k) e^{i\omega(t-r)}\,.
\end{eqnarray}
The component $\epsilon_-^\mu j_\mu (k)$ is nothing but the EM current projected onto a helicity state. It is related to the quantum current, $J^{\mu}$ by our classical limit prescription. Following LSZ reduction we can identify the latter with the scattering amplitude for the process under consideration,
\begin{equation}\label{eq:clasamps1}
    \epsilon_-^\mu j_\mu (k) = \epsilon_-^\mu \langle J_\mu (k)\rangle   =\langle A_4^{h=-1}\rangle \,.
\end{equation}
In order to match \eqref{eq:psi2rad} with the result of the Teukoslky equation we need to perform an expansion into \textit{spin-weighted} spherical harmonics carrying helicity $h=-1$ under the transformation \eqref{eq:heltr}. This will be relegated to Part II, where we will provide a simple representation of such harmonics together with a projection method.

We are now in good position to evaluate the classical amplitude \eqref{eq:clasamps1}. First, by choosing  $\phi=0$ in our null tetrad \eqref{eq:nlltet} we note that the vector $l$ for the scattered wave is precisely aligned with $k_3$ in \eqref{eq:kinematics}. Thus we identify $\theta$ as the scattering angle. The corresponding polarization directions are read off from \eqref{eq:nlltet}:

\begin{equation}\label{eq:photon polarization1}
    \begin{split}
        \epsilon_{3}^{+}= & m=\frac{1}{\sqrt{2}}(0,\cos\theta, i,-\sin\theta)\, ,\\
\epsilon_{3}^{-}  =&
-\bar{m}=- \frac{1}{\sqrt{2}}(0,\cos\theta,- i,-\sin\theta)\,. 
    \end{split}
\end{equation}
An analogous construction holds for the incoming wave. We set the incoming momentum $k_2$ in \eqref{eq:kinematics} as corresponding to the same configuration with $\theta=0$, namely 

\begin{equation}\label{eq:photon polarization2}
    \begin{split}
    k_2= & (1,0, 0,1) \,,\\
        \epsilon_{2}^{+}= & \frac{1}{\sqrt{2}}(0,1, i,0) \,,\\
\epsilon_{2}^{-} =& - \frac{1}{\sqrt{2}}(0,1,- i,0)\,. 
    \end{split}
\end{equation}
To evaluate the graviton exchange diagram we will need the three-vertex of two photons legs and one graviton. In the harmonic gauge it reads
 \cite{Holstein:2008sx}:
 \begin{equation}
\begin{split}
V_{\gamma\gamma h}^{\alpha,\beta,\mu\nu} & =-\frac{i\kappa}{2}\left[\eta^{\mu\nu}\left(k_{2}{\cdot}k_{3}\eta^{\alpha\beta}-k_{2}^{\beta}k_{3}^{\alpha}\right)-2I_{\,\,\,\,\kappa\lambda}^{\mu\nu}\left(k_{2}{\cdot}k_{3}I^{\alpha\beta,\kappa\lambda}+k_{2}^{(\kappa}k_{3}^{\lambda)}\right)\eta^{\alpha\beta}\right. \\
 & \,\,\,\,\,\,\,\,\,\,\,\,\,\,\,\,\,\,\,\,\,\,\,\,\,\,\,\left.-\left(k_{2}^{\kappa}k_{3}^{\alpha}\eta^{\lambda\beta}+k_{3}^{\kappa}k_{2}^{\beta}\eta^{\lambda\alpha}\right)\right],\,\quad   I^{\alpha\beta,\gamma\delta}:=\frac{1}{2}\left(\eta^{\alpha\gamma}\eta^{\beta\delta}+\eta^{\alpha\delta}\eta^{\beta\gamma s}\right).  \label{eq:vertex photon}
\end{split}
\end{equation}
The tree-level scattering amplitude is then obtained by gluing this vertex to  \eqref{eq:vertex scalar scalar graviton}. It becomes 
\begin{equation}
    \begin{split}
      A_{4}^{h=1} & =-\frac{\kappa^{2}}{4q^{2}}\left[p_{1}{\cdot}k_{2}\left(4p_{1}{\cdot}k_{2}+2q^{2}\right)\epsilon_{2}{\cdot}\epsilon_{3}^{*}+\left(2k_{2}{\cdot}\epsilon_{3}^{*}q{\cdot}\epsilon_{2}{+}q^{2}\epsilon_{2}{\cdot}\epsilon_{3}^{*}\right)M^{2}\right.\\
 & \,\,\,\,\,\,\,\,\,\,\,\,\,\,\,\,\,\,\,\,\,\,\,\left.-2q^{2}p_{1}{\cdot}\epsilon_{2}\left(p_{1}{+}k_{2}\right){\cdot}\epsilon_{3}^{*}-4p_{1}{\cdot}k_{2}\left(p_{1}{\cdot}\epsilon_{2}k_{2}{\cdot}\epsilon_{3}^{*}{+}p_{1}{\cdot}\epsilon_{3}^{*}q{\cdot}\epsilon_{2}\right)\right],  
    \end{split}
\end{equation}
where $2p_1\cdot k_2=(s-M^2),q^2=t$. As explained, in the classical limit we need to keep all the terms  that scale as
$\omega^{0}$. This means that the $q^2$ terms in the numerator, i.e. contact terms, will indeed contribute at the same footing as the t-channel residue. The full classical result is then
\begin{equation}
\label{photon-SBH}
    \begin{split}
\langle A_{4}^{h=1}\rangle & =-\frac{\kappa^{2}}{4q^{2}}\left[\left(2p_{1}{\cdot}k_{2}\right)^{2}\epsilon_{2}{\cdot}\epsilon_{3}^{*}+\left(2k_{2}{\cdot}\epsilon_{3}^{*}q{\cdot}\epsilon_{2}{+}q^{2}\epsilon_{2}{\cdot}\epsilon_{3}^{*}\right)M^{2}\right.\\
 & \,\,\,\,\,\,\,\,\,\,\,\,\,\,\,\,\,\,\,\,\,\,\,\left.-2q^{2}p_{1}{\cdot}\epsilon_{2}p_{1}{\cdot}\epsilon_{3}^{*}-4p_{1}{\cdot}k_{2}\left(p_{1}{\cdot}\epsilon_{2}k_{2}{\cdot}\epsilon_{3}^{*}{+}p_{1}{\cdot}\epsilon_{3}^{*}q{\cdot}\epsilon_{2}\right)\right]. 
    \end{split}
\end{equation}
We now consider different helicity combinations, thereby evaluating the elements of the scattering matrix \eqref{eq:scattering-matrix-helicities}. For the helicity preserving combination,
\begin{equation}\label{eq:photona4}
    \langle A_{4,++}^{h=1}\rangle=\langle A_{4,--}^{h=1}\rangle=\frac{\kappa^{2}M^{2}}{4}\frac{\cos^{2}\left(\frac{\theta}{2}\right)}{\sin^{2}\left(\frac{\theta}{2}\right)}\,.
\end{equation}
On the other hand, we find that in the classical limit the helicity reversing amplitudes, $  \langle A_{4,+-}^{h=1}\rangle,\langle A_{4,-+}^{h=1}\rangle$, vanish. This is consistent with the intuition that in the soft limit, photons $k_2,k_3$ behave as a matter line of definite chirality, which sources a classical stress-energy tensor (here represented by the exchanged soft graviton \cite{Weinberg:1965nx}). At subleading order $\mathcal{O}(\hbar \omega) $ such soft behaviour is modified and the helicity of the photon can flip. This picture is true as long as $h<2$, as for $h=2$ there emerge different Feynman diagrams for graviton interactions leading to a helicity flip even at the classical level $(\hbar \omega)^0$, as we will  see in \cref{sec:gw-scattering-schw}. 

To confirm the above picture about conservation of helicity, we briefly discuss here the scattering of neutrino ($\pm h=\pm 1/2$) waves \cite{futterman88}. For this we translate \eqref{eq:nlltet} into spinor-helicity variables, see \cite{Elvang:2013cua} for conventions. Using  $v_{\alpha \dot{\alpha}}=\sigma^\mu_{\alpha \dot{\alpha}} v_\mu$ to convert Lorentz vectors $v^\mu$ into spinors,   we obtain 

\begin{eqnarray}\label{eq:vectospin}
     k_{\alpha \dot{\alpha}} = \omega  l_{\alpha \dot{\alpha}} = |\lambda\rangle_{\alpha} [\tilde{\lambda}|_{\dot \alpha}\, &,&  n_{\alpha \dot{\alpha}}=  |\mu \rangle_{\alpha} [\tilde{\mu}|_{\dot \alpha} \,, \nonumber \\
     m_{\alpha \dot{\alpha}} =   \epsilon^{+}_{\alpha \dot{\alpha}} = \sqrt{2} \frac{|\mu \rangle_{\alpha} [\tilde{\lambda}|_{\dot \alpha}}{\langle \lambda \mu\rangle }\, &,&  \bar{m}_{\alpha \dot{\alpha}}=-\epsilon^{-}_{\alpha \dot{\alpha}}=  \sqrt{2}\frac{|\lambda \rangle_{\alpha} [\tilde{\mu}|_{\dot \alpha} }{[\tilde{\mu} \tilde{\lambda}]}   \,.
\end{eqnarray}
The relation $l\cdot n = -m\cdot \bar m =1$ now becomes $|\langle \lambda \mu\rangle |= |[\tilde{\lambda}\tilde{\mu}]|=\sqrt{2\omega}$. The Weyl spinors $|\lambda\rangle$, $|\lambda]$ carry helicity weight $-1/2,+1/2$ under \eqref{eq:heltr}, respectively.\footnote{The spinorial description of spacetime \eqref{eq:gnullex} has been developed in \cite{Penrose:1987uia}, and this parametrization will play a crucial role for GW scattering in Part II of this work. Spinors have been recently used in \cite{Chacon:2021wbr} to expand the classical double copy construction  \cite{Monteiro:2014cda, Monteiro:2015bna}.} Fixing the little group we can write them explicitly as 
\begin{equation}\label{eq:spinhel}
    \begin{split}
      |\mu\rangle&= \left(e^{-i\phi/2} \cos\theta/2, e^{i\phi/2} \sin\theta/2 \right),\,\,\, |\lambda\rangle = \sqrt{2\omega}\left(-e^{-i\phi/2} \sin\theta/2, e^{i\phi/2} \cos\theta/2 \right),\\
      [\tilde{\mu}| &=\left(e^{i\phi/2} \cos\theta/2, e^{-i\phi/2} \sin\theta/2 \right),\,\,\, [\tilde{\lambda}| = \sqrt{2\omega}\left(-e^{i\phi/2} \sin\theta/2, e^{-i\phi/2} \cos\theta/2 \right).
    \end{split}
\end{equation}
Note that, nicely, these spinors can be used to construct both solutions of the massless Dirac equation for momentum $k=\omega l$:

\begin{equation}\label{eq:dirac-spinors}
    u^-(k)= \left(|\lambda\rangle_{\alpha},0,0 \right)^T \,, \,\,u^+(k)= \left(0,0,|\tilde{\lambda}]^{\dot{\alpha}}\right)^T\,.
\end{equation}
The states associated to $k_3,k_2$ are again recovered by setting $\phi=0$ (and $\theta=0$ for the latter) in \eqref{eq:spinhel}. Plugging these into the corresponding four-point graviton exchange amplitude, which is given by \cite{Holstein:2008sx}
\begin{equation}
   A_{4}^{h=\frac{1}{2}}=\frac{\kappa^2}{8q^{2}}\left[4p_{1}{\cdot}k_{2}+q^{2}\right]\bar{u}_{3}\slashed{p}_{1}u_{2}\,,
\end{equation}
and taking the classical limit, we obtain
\begin{equation}\label{eq:spinora4}
  \langle A_{4,++}^{h=\frac{1}{2}}\rangle=\langle A_{4,--}^{h=\frac{1}{2}}\rangle=-\frac{\kappa^{2}M^{2}}{4}\frac{\cos\left(\frac{\theta}{2}\right)}{\sin^{2}\left(\frac{\theta}{2}\right)}\,,
\end{equation}
whereas the helicity-reversing elements vanish as expected.

Finally, collecting the results \eqref{eq:crsscalar},\eqref{eq:photona4},\eqref{eq:spinora4}, the classical cross-sections \eqref{eq:classical cross section} can be written in the compact formula 

\begin{equation}\label{eq:crforh}
  \boxed{  \frac{d\langle \sigma^h\rangle}{d\Omega}= G^2 M^2\frac{\cos^{4h}\left(\theta/2\right)}{\sin^4\left(\theta/2\right)}}\,,
\end{equation}
for $h=0,1/2,1$, in perfect agreement with the results in \cite{PhysRevD.16.237,Westervelt:1971pm,Doran:2001ag,Dolan:2007ut}. Of course, this result is physical and independent of our choice of little-group fixing. An interesting `double copy' structure is manifest, with the numerator for $h=1$ being the square of the $h=1/2$ case. This can be easily seen at the level of the amplitudes by plugging spinor-helicity variables \eqref{eq:vectospin} into the photon result \eqref{photon-SBH}. We now show how the pattern in \eqref{eq:crforh} breaks down for $h=2$ due to the fact that the helicity-violating amplitudes do not vanish.

\subsection{Gravitational Waves}\label{sec:gw-scattering-schw}

Even though the case of GW will be extensively covered in Part II for the Kerr background, let us here present the classical prescription leading to GW scattering in Schwarzschild. We will follow an analogous route and relate the graviton Compton amplitude $A_4^{h=2}$ to the corresponding NP scalar. In this case the Weyl scalar reads \cite{Newman_1962}

\begin{equation}\label{eq:psi2gr}
    \boldsymbol{\Psi}^{\textrm{rad}}_{-2}=- \bar{m}^\mu n^\nu \bar{m}^{\rho} n^\sigma C^{\textrm{rad}}_{\mu \nu \rho \sigma}\,.
\end{equation}
Here the radiative piece of the Weyl tensor is obtained from a linearized metric perturbation sourced by the  on-shell modes of the stress-energy tensor

\begin{equation}
    h_{\mu \nu}^{\textrm{rad}} \propto T_{\mu \nu}(k) e^{i k\cdot x}\,,
\end{equation}
with $k^2=0$. The conservation condition $k^\mu T_{\mu \nu}=0$ is equivalent, through the gauge fixing \eqref{eq:harmgauge}, to the fact that the metric perturbation is transverse-traceless (TT):

\begin{equation}
    \partial^\mu h_{\mu\nu}^{\textrm{rad}}=\eta^{\mu \nu}h_{\mu\nu}^{\textrm{rad}}=0\,.
\end{equation}
We use this to conclude that all contractions of the linearized Riemann tensor
\begin{equation}
    R^{\textrm{rad}}_{\alpha \beta \mu \nu}=-\frac{1}{2}(\partial_{\alpha \mu }h^{\textrm{rad}}_{\beta\nu}+\partial_{\beta\nu}h^{\textrm{rad}}_{\alpha\mu}-\partial_{\alpha\nu}h^{\textrm{rad}}_{\beta\mu}-\partial_{\beta \mu}h^{\textrm{rad}}_{\alpha \nu})\, ,
\end{equation}
vanish. This means that $C^{\textrm{rad}}_{\alpha \beta \mu \nu}=R^{\textrm{rad}}_{\alpha \beta \mu \nu}$ for the radiative modes. The Weyl NP scalar \eqref{eq:psi2gr} then becomes

\begin{equation}
 \boldsymbol{\Psi}^{\textrm{rad}}_{-2}\propto \frac{1}{2} \bar{m}^\alpha n^\beta \bar{m}^{\mu} n^\nu \left(k_{\alpha} k_{ \mu }T_{\beta\nu}+k_{\beta}k_{\nu}T_{\alpha\mu}-2k_{\alpha}k_{\nu}T_{\beta\mu}\right)e^{i k\cdot x}\, ,
\end{equation}
and using the saddle-point approximation \eqref{eq:saddle} we obtain

\begin{equation}
 \boldsymbol{\Psi}^{\textrm{rad}}_{-2}\propto \frac{\omega^2}{2} \bar{m}^\mu \bar{m}^{\nu}  T_{\mu\nu}(k)e^{i \omega (t-r)}\,.
\end{equation}
Identifying the graviton polarizations
\begin{equation}\label{eq:grpol}
    \epsilon^{\mu\nu}_{+} =\epsilon_{+}^\mu \epsilon_{+}^\nu =  m^\mu m^{\nu}\,,\,\,  \epsilon^{\mu\nu}_{-} =\epsilon_{-}^\mu \epsilon_{-}^\nu =  \bar{m}^\mu \bar{m}^{\nu}\,,
\end{equation}
we can finally relate the NP scalar to the classical limit of the scattering amplitude 
\begin{equation}
    \bar{m}^\mu \bar{m}^{\nu}  T_{\mu\nu}(k)=\langle A_4^{h=2}\rangle\,.
\end{equation}

In order to evaluate this let us note a crucial difference with previous cases. The classical
amplitude $\langle A_4^{h=2}\rangle$ is controlled not only by the single graviton exchange diagram,
but actually, all three channels contribute. The additional $s$ and $u$ channels open up due to the response of the BH to the gravitational wave perturbation.  Using the factorized states \eqref{eq:grpol}, the amplitude can be written as the double
copy of the scalar-QED Compton amplitude \cite{Bjerrum-Bohr:2013bxa}
\begin{equation}\label{eq:compton0}
  A_{4}^{h=2}=-\kappa^{2}\frac{\left(2p_{1}{\cdot}F_{2}{\cdot}F_{3}{\cdot}p_{1}\right)^{2}}{\left(s-M^{2}\right)\left(u-M^{2}\right)t}\,,  
\end{equation}
 where $F_{i}^{\mu\nu}=2k_{i}^{[\mu}\epsilon_{i}^{\nu]}$.

The gravitational amplitude contains poles corresponding to $s,u$ and $t$-channels. The residues of these get mixed under gauge transformations of the external gravitons. Moreover, under the classical limit \eqref{eq:sandtclassical} and finite scattering angle $\xi$, we learn that the $s$ and $u$ channels form a double pole in $s-M^2\sim M^2-u$, with the same classical scaling as $t\sim(\hbar \omega)^2$. This means that the contributions from all diagrams are entangled in the classical limit, in contrast with previous gauge-dependent approaches \cite{Guadagnini:2008ha,Westervelt:1971pm}.\footnote{Indeed, it was argued in  \cite{Guadagnini:2008ha,Westervelt:1971pm} that the classical amplitude for the scattering of gravitational waves off classical matter has only the contribution of the graviton exchange diagram. However, we have checked that this fact turns out to be true only in the spatial gauge $\epsilon^{0\mu}_i=0$ of the gravitons. This trick also breaks down for the rotating case, so we choose instead to simply work in a manifestly Lorentz covariant gauge.}

 Using the kinematics $(\ref{eq:kinematics})$, and the basis for
polarization vectors  $(\ref{eq:photon polarization1}),(\ref{eq:photon polarization2})$
it is now easy to show that the diagonal elements of the classical scattering matrix \eqref{eq:scattering-matrix-helicities} 
are given by 
\begin{equation}
\langle A_{4,++}^{h=2}\rangle=\langle A_{4,--}^{h=2}\rangle=\frac{\kappa^{2}}{4}\frac{\cos^{4}\left(\frac{\theta}{2}\right)}{\sin^{2}\left(\frac{\theta}{2}\right)}\,,\label{eq:m+-gravitons}
\end{equation}
whereas for the off-diagonal elements we have 

\begin{equation}
\langle A_{4,+-}^{h=2}\rangle=\langle A_{4,-+}^{h=2}\rangle=\frac{\kappa^{2}}{4}\frac{\sin^{4}\left(\frac{\theta}{2}\right)}{\sin^{2}\left(\frac{\theta}{2}\right)}\,.\label{eq:m4++gravitons}
\end{equation}
We therefore conclude that helicity is \textit{not} preserved for GW scattering. For the case of  an incoming gravitational
wave with fixed positive helicity, the unpolarized differential cross section  \eqref{eq:classical cross section} becomes \footnote{Since  \eqref{eq:compton0} is a double copy amplitude, it is easy to check that for  single copy case, the classical  differential cross section is given by  $\frac{d\langle\sigma^{\rm{QED}}\rangle }{d\Omega}= \left(\frac{e^2}{4\pi M}\right)^2\left[\cos^4\left(\frac{\theta}{2}\right)+\sin^4\left(\frac{\theta}{2}\right)\right]$, which recovers the unpolarized differential cross section for the Thompson scattering \cite{Jackson:100964}. }

\begin{equation}\label{eq:crgr}
   \frac{d\langle \sigma^{h=2}\rangle}{d\Omega}=\frac{G^2M^2}{\sin^{4}\left(\frac{\theta}{2}\right)}\left[\cos^{8}\left(\frac{\theta}{2}\right)+\sin{}^{8}\left(\frac{\theta}{2}\right)\right]
\end{equation}
in agreement with \cite{PhysRevD.16.237,Westervelt:1971pm,Doran:2001ag,Dolan:2007ut}. Then as 
announced above, the GW scattering cross section differs from the lower-helicity result \eqref{eq:crforh}. There is an extra sine term in the unpolarized differential
cross section for $h=2$ arising from
the opposite helicities configuration of the Gravitational Waves \eqref{eq:m4++gravitons}. In the black hole perturbation theory framework \cite{Dolan:2008kf}, this was explained from the parity dependence of the GW-BH interaction as we will outline and expand in Part II.

This new term does not alter the universal $\theta \to 0$ Newtonian behaviour \eqref{eq:newtonian}. Moreover, we will confirm that GW scattering leads to the same eikonal behaviour as the lower-helicity waves in \cref{Eikonal-app}.

To close this section, let us discuss the effect of the BH on the polarization of the scattered wave. We note that the cross-section \eqref{eq:crgr}, as well as the previous examples, does not depend on the helicity $\pm h$ of the incoming wave. This means that for a given scattering angle $\theta$ in a detector, left and right polarized beams will scatter with the same probability and the average polarization is preserved. This does not hold if the black hole is spinning as one helicity will be favoured, as we show in the next section.

\section{Massless wave scattering off a Kerr background}\label{sec:Section3}

In this section we will extend the previous framework to study the scattering of massless
plane waves off the Kerr BH. In the 1970's, motivated by the stability problem of Kerr, preliminary analyses found that even for slowly spinning bodies there are interesting observable effects. For instance, the spin of the black hole can induce a partial polarization of electromagnetic or gravitational waves after the scattering process \cite{PhysRevD.16.237, Dolan:2008kf}. We will review and extend these results. 

Our tool will be the relation between scattering of spinning black holes and higher-spin amplitudes. A connection between classical rotating bodies and massive particles with spin
was first pointed out in \cite{Vaidya:2014kza} and extended in \cite{Guevara:2017csg}. It was suggested that QFT amplitudes of massive particle of spin $S$ emitting gravitons encode the gravitational coupling to $2S$ classical multipoles, namely the powers $a^j$ for $j\leq 2S$. The main evidence for this comes from the  multipoles of a single Kerr black hole metric, which are encoded in the on-shell modes of its effective stress-energy tensor \cite{Vines:2017hyw}, and can be mapped to three-point amplitudes by introducing the QFT notion of \textit{minimal-coupling}. This correspondence was tested by studying the scattering of two black holes to leading order in $G$ \cite{Cachazo:2017jef,Guevara:2017csg,Guevara:2018wpp,Guevara:2019fsj,Maybee:2019jus,Chung:2018kqs,Chung:2019duq,Chung:2020rrz}

In the previous section, wave scattering in the Schwarzschild background was modeled using the off-shell three-point vertex \eqref{eq:vertex scalar scalar graviton} for a quantum massive scalar coupled to gravity, uniquely defined assuming minimal coupling and the harmonic gauge. For spinning particles these considerations are not enough: On the one hand, there is no clear notion of minimal coupling at four-points, so even though three-point amplitudes for Kerr can indeed be fixed by minimal coupling as in \cite{Guevara:2017csg}, this prescription breaks down for four-point amplitudes \cite{Ferrara:1992yc}.\footnote{At the quantum level, massive higher-spin particles can exhibit diverse inconsistencies related to its bad UV behaviour \cite{Ferrara:1992yc,Cucchieri:1994tx,Arkani-Hamed:2017jhn}. Even though these are tamed at three-points by introducing certain criteria for minimal coupling \cite{Deser:2000dz,Deser:2001dt,Arkani-Hamed:2017jhn}, reflecting the fact that their massless limit is still consistent at three-points \cite{Benincasa:2007xk}, they do persist at four-points leading to acausality/superluminality \cite{Deser:2001dt,Afkhami-Jeddi:2018apj,Camanho:2014apa,Bautista:2019evw}.} On the other hand, given that four-point amplitudes can be constructed from Feynman diagrams we may be tempted to extend the aforementioned three-point amplitude to an (off-shell) vertex for a higher-spin quantum particle. Such an off-shell extension  is however not unique even assuming minimal coupling and carries the ambiguities of the multipole decomposition of higher-spin objects, see e.g \cite{Lorce:2009br} and subsequent work.

In this section we extend the correspondence to the four point amplitudes representing the scattering of a massless particle/wave off the the Kerr black hole, as an infinite-spin massive particle. In order to sort out the above complications at four-points we will instead take a more indirect approach. We assume the existence of a QFT three-point vertex, containing all multipoles $\sim a^j$, such that in the classical limit it reduces to the off-shell Kerr metric. More precisely, assuming the harmonic gauge, the classical limit of the vertex should be equivalent to the energy-momentum tensor presented in a compact form in \cite{Vines:2017hyw}. This extends to all spin-multipoles the seminal approach of \cite{PhysRevD.16.237} to wave scattering, which used a background gravitational field containing a spin monopole/dipole as source. Also, by extending the classical rules given in the previous section, we will check that our infiite-spin vertex indeed has classical scaling and can be used for our amplitudes.

Let us remark that in this part of the work we will use the described Feynman rules to compute the amplitudes for $h<2$. These amplitudes inherit the exponential structure of the three-point amplitude and hence are free of high-energy divergences $M\to 0$. In Part II, however, we will see that the GW case ($h=2$) indeed contain such inconsistencies and can only be constructed via an on-shell ansatz since there are no natural choice of Feynman rules for couplings $ R^2 \sim G^2 $.

\subsection{Classical Expansion}

To begin establishing the correspondence we first define the classical variables into play. It was explained in \cref{sec:2} that the classical limit is equivalent to the scaling $r\to \hbar^{-1} r$, for a scattering radius (i.e. impact parameter) $r$. This is equivalent to an expansion in large orbital angular momentum, $L \sim r \times p$,

\begin{equation}\label{eq:clasL}
    L = \hbar^{-1} \hat{L}\,.
\end{equation}
This perspective has been detailed in e.g. \cite{Damour:2017zjx,Cheung:2018wkq}. We can now extend the scaling rule to the intrinsic part of the angular momentum. Let $a$ be the norm of the Pauli-Lubanski vector of a massive particle. Then, the intrinsic angular momentum is given by $S=M a$ and the rule \eqref{eq:clasL} is extended by considering \cite{Maybee:2019jus}

\begin{equation}\label{eq:classical-limit-spin}
    a =  \hbar^{-1} \hat{a}\,.
\end{equation}
The classical variable $\hat{a}$ can be identified with the radius of the ring-singularity in Kerr \cite{Guevara:2018wpp}. As a consequence of this scaling, the expansions considered in Section \ref{sec:2} can be extended by an expansion in the following dimensionless quantities:

\begin{equation}
  \textcolor{black}{a^\star}  = \frac{a}{GM} \,  ,\quad a\omega  \,,
\end{equation}
where we recall that $\omega=\hbar \hat{\omega}$ is the energy of the massless momenta. Both dimensionless combinations scale as $\hbar^0$ and hence are classical (we know that e.g. $a\omega=\hat{a}\hat{\omega}$, thus the hat notation can be omitted hereafter). As their ratio is given by $a\omega/\textcolor{black}{a^\star}=GM\omega = \epsilon/2$, corresponding to our classical weak-field parameter, only one new expansion is needed to incorporate the full spin dependence. As we are working with tree-level amplitudes at leading order in $G$, it is more natural to perform an expansion in $a\omega <1 $. It is easy to see that this corresponds to a soft expansion in the external massless momenta, as advocated in \cite{Guevara:2018wpp,Bautista:2019evw}. The expansion involves corrections in $a\omega \sim 1/\lambda$ that can probe the internal structure of the spinning body, although as explained the orders $(a\omega)^0$ and $(a\omega)^1$ correspond to spin monopole and dipole respectively, and are universal. More generally, the $2S+1$ classical spin multipoles of the scattering amplitude $\langle A^{S,h}_4 \rangle$, for a spin-S particle, are in one-to-one correspondence with the soft expansion up to order $(a\omega)^{2S}$.

\subsection{Scattering of waves of helicity $h<2$}

As anticipated, for lower helicity waves we will \textit{not} model the Kerr BH by a three point vertex between a massive quantum particle and a graviton, which was our approach in the previous section. Instead, we will directly consider the off-shell classical source given in \cite{Vines:2017hyw}. Motivated by the approach in \cite{PhysRevD.16.237} we will use this source as a background field to compute the scattering amplitude with a wave of helicity $h$, denoted by $\langle A_4^{S\to \infty ,h} \rangle$. It is given by a single graviton exchange diagram with a massless particle of helicity $h$, where the internal graviton is required to be off-shell in order to capture the contact terms at finite scattering angle $\theta$. Here $S\to \infty$ means that we are considering the full infinite series in $(a \omega)$, as opposed to the linear in spin results of \cite{PhysRevD.16.237}.

To begin with, consider the linearized form of the Kerr metric as in  \cite{Vines:2017hyw,Harte:2016vwo}, satisfying the harmonic gauge condition \eqref{eq:harmgauge}\footnote{This form of the vertex differs by ultralocal pieces, e.g. $\delta(r)$, from the compact expression (31) given in \cite{Vines:2017hyw}. This pieces will lead to $\mathcal{O}(q^0)$ corrections in eq. \eqref{eq:justin vertex classical}, which were irrelevant in \cite{Vines:2017hyw} but do contribute to our amplitude at finite scattering angle. We consider the form in \eqref{eq:justin position } to be accurate at leading order in $G$ as it follows essentially from the Newman-Janis shift \cite{Harte:2016vwo}.}

\begin{equation}
    h_{\mu\nu}(x)=\mathcal{P}_{\mu\nu\alpha\beta}\left(u^{\alpha}u^{\beta}\cos{a{\cdot}\partial}{+}\frac{\sin{a{\cdot}\partial}}{a{\cdot}\partial}u^{(\alpha}\epsilon_{\,\,\rho\sigma\delta}^{\beta)}u^{\rho}a^{\sigma}\partial^{\delta} \right)\frac{4GM}{r},\label{eq:justin position }
\end{equation}
where $u$ describes the four-velocity of the black-hole, in the rest frame $u=(1,0,0,0)$. In general we will identify $u=p_1/M$. As we further elaborate below, its spin is encoded in the constant pseudovector $a^\mu$ satisfying $a\cdot u=0$. Noting that $\partial\sim 1/r \sim \omega$ we note that this form corresponds to an infinite expansion in $a\omega$, while only considering the leading order in $GM/r\sim GM\omega$ as desired (recall that both combinations are classical). To construct the scattering amplitude we map this form to momentum space writing the retarded green function as  \cite{Vines:2017hyw}
\begin{equation}
    \frac{1}{ 4\pi r} := \int d\tau  G_{ret}(x,u\tau)= \int d\tau \int \frac{d^4 q}{(2\pi)^4q^2 }  e^{iq\cdot (x-u\tau)} =  \int \frac{d^4 q}{(2\pi)^4q^2 }  (4\pi)\delta(2q\cdot u )e^{iq\cdot x}\,.
\end{equation}
where we have ignored the contour ($i\epsilon$) prescription. Plugging this into \eqref{eq:justin position } we can read off, up to a  normalization, the three-point off-shell vertex corresponding to the Kerr background in momentum space:

\begin{equation}
\tilde{h}_{\mu\nu}(q)=\frac{\kappa}{q^{2}}\delta(2q\cdot p_1)\mathcal{P}_{\mu\nu\alpha\beta}\left(p_{1}^{\alpha}p_{1}^{\beta}\cosh{a{\cdot}q}{+}i \frac{\sinh{a{\cdot}q}}{a{\cdot}q}p_{1}^{(\alpha}\epsilon_{\,\,\rho\sigma\delta}^{\beta)}p_{1}^{\rho}a^{\sigma}q^{\delta} \right),\label{eq:justin vertex classical}
\end{equation}
where $p_1^2=M^2$, and we note that $\delta(2q\cdot p_1)$ is nothing but the classical limit of the delta function $\delta(p_4^2-m^2=2q\cdot p_1 + q^2)$,which will be omitted in our scattering amplitude. Note also that the graviton momentum $q^\mu$ is still off-shell.

Regarded as a Feynman vertex, the form \eqref{eq:justin vertex classical} extends to all orders in spin the scalar vertex \eqref{eq:vertex scalar scalar graviton} including graviton propagator \eqref{eq:propagator graviton} (note however that the classical limit is not required as this is a background field and not a quantum particle). The scalar piece, i.e. `spin monopole', corresponds to the first term $p_1^\alpha p_1^\beta$, whereas the universal spin dipole is given by the coupling $p_{1}^{(\alpha}\epsilon_{\,\,\rho\sigma\delta}^{\beta)}p_{1}^{\rho}a^{\sigma}q^{\delta}$. The remaining multipoles $\sim (a\cdot q)^n$ are fixed completely by the exponential structures.

In order to compute the scattering amplitudes we will follow the same logic as in the previous section. Indeed, the argument leading to e.g. \eqref{eq:psiphoton}-\eqref{eq:psi2rad}, relating the radiative part of the Weyl scalar $\boldsymbol{\Psi}^h$ to the classical amplitude $\langle A^h_4\rangle$ holds irrespective of the scattering source and spin effects.

Using the variables given in \eqref{eq:kinematics}, let us further choose a generic  orientation for  the spin of the black hole,
\begin{equation}
a^{\mu}=(0,a_x,a_y,a_z).\label{eq:polar spin}
\end{equation}
The computation of $\langle A_4^h\rangle$ for waves
of helicity $h<2$ follows from a single graviton-exchange diagram as in the previous section. We use the vertex $(\ref{eq:justin vertex classical})$
for the massive line and the Feynman rules from \cite{Holstein:2008sx} for the massless particles.

\textit{Scalar scattering}. For the scattering of scalar waves, the classical amplitude reads
\begin{equation}\label{scalar-kerr}
\langle A_{4}^{S\to \infty ,h=0} \rangle=\frac{\kappa^2 p_1{ \cdot} k_2}{ q^2}\left[p_1{ \cdot} k_2 \cosh{a{\cdot}q}+ i M \vec{a}{\cdot}\vec{q}{\times}\vec{k}_2 \frac{\sinh{a{\cdot}q}}{a{\cdot}q} \right].    
\end{equation}
 After using the kinematics  $(\ref{eq:kinematics})$ and $(\ref{eq:polar spin})$ this amplitude takes the form
\begin{equation}\label{eq:scalar_kerr}
\langle A_4^{S\to \infty,h=0}\rangle=\frac{M^2\kappa^2 }{ 4\sin^2\left(\frac{\theta}{2}\right)}\left[ \cosh{a{\cdot}q}- i\omega a_y\sin\theta  \frac{\sinh{a{\cdot}q}}{a{\cdot}q} \right],       
\end{equation}
where $q{\cdot}a= -\omega( a_x\sin(\theta)-2 a_z\sin^2(\theta/2))$.

\textit{Neutrino scattering}. For the scattering of neutrino waves it follows that

\begin{equation}\label{neutrino-kerr}
    \langle A_{4}^{S\to \infty,h=1/2} \rangle{=}\frac{\kappa^2 }{2 q^2}\bar{u}_3\left[p_1{ \cdot} k_2\slashed{p}_1\cosh{a{\cdot}q}{-} iM\left(p_1{ \cdot} k_2\vec{a}{\cdot}\vec{\gamma}{\times}\vec{q}{-}\vec{a}{\cdot}\vec{q}{\times}\vec{k}_2\slashed{p}_1\right)\frac{\sinh{a{\cdot}q}}{2a{\cdot}q} \right]u_2,
\end{equation}
and in terms of the kinematics $(\ref{eq:kinematics})$, the basis of spinors \eqref{eq:dirac-spinors}, and the spin vector $(\ref{eq:polar spin})$, the diagonal elements of the scattering matrix \eqref{eq:scattering-matrix-helicities} read
\begin{equation}\label{photon-kerr}
\begin{split}
     \langle A_{4,++}^{S\to\infty,h=1/2} \rangle&=\frac{M^2\kappa^2 \cos\left(\frac{\theta}{2}\right)}{4\sin^2\left(\frac{\theta}{2}\right)}\biggl[\cosh{a{\cdot}q}{-}\omega \tan\left(\theta/2\right)\times\\
&    \hspace{1.7cm}\left(2a_x\sin^2\left(\theta/2\right){+}2i a_y\left(1{+}\cos^2\left(\theta/2\right)\right)
     {+}a_z\sin\theta\right)\frac{\sinh{a{\cdot}q}}{2a{\cdot}q}\bigg]\\
     \langle A_{4,--}^{S\to\infty,h=1/2} \rangle &= \left[\langle A_{4,++}^{\rm{K},h=1/2}\rangle^* \right]_{\omega\rightarrow-\omega},
     \end{split}
\end{equation}
whereas as for the spin-less case,  the off-diagonal elements vanish.
 
 \textit{Photon scattering}. Finally, for the scattering of electromagnetic waves, the classical amplitude is given by
\begin{equation}
    \langle A_4^{S\to \infty,h=1}\rangle{=}\langle A_4^{h=1}\rangle\cosh{a{\cdot}q}{+}\frac{ i M \kappa^2}{2q^2}\left[2\epsilon_2{\cdot}\epsilon_3^*\vec{a}{\cdot}\vec{q}{\times}\vec{k}_2{-}\epsilon_3^*{\cdot}k_2\vec{a}{\cdot}\vec{q}{\times}\vec{\epsilon}_2{-}\epsilon_2{\cdot}k_2\vec{a}{\cdot}\vec{q}{\times}\vec{\epsilon}_3^*\right]\frac{\sinh{a{\cdot}q}}{a{\cdot}q},
\end{equation}
where $\langle A_4^{h=1}\rangle$ is the classical amplitude for the scattering of electromagnetic waves off the  SBH, given in \eqref{photon-SBH}. For this case, the elements of \eqref{eq:scattering-matrix-helicities} are 
\begin{equation}\label{eq:pppht}
\begin{split}
     \langle A_{4,++}^{S\to \infty,h=1} \rangle&=\frac{M^2\kappa^2 \cos^2\left(\frac{\theta}{2}\right)}{4\sin^2\left(\frac{\theta}{2}\right)}\biggl[\cosh{a{\cdot}q}{-}\omega \tan\left(\theta/2\right)\left(2a_x\sin^2\left(\theta/2\right){+}i 2 a_y
     {+}a_z\sin\theta\right)\frac{\sinh{a{\cdot}q}}{a{\cdot}q}\bigg],\\
     \langle A_{4,--}^{S\to\infty,h=1} \rangle &= \left[\langle A_{4,++}^{S\to\infty,h=1} \rangle^*\right]_{\omega\rightarrow-\omega},
     \end{split}
\end{equation}
and similar to the previous case, the off-diagonal elements vanish.

We have checked that to linear order in spin, amplitudes \eqref{scalar-kerr}, \eqref{neutrino-kerr} and \eqref{photon-kerr} recover the result for the scattering of scalar, neutrino and electromagnetic waves off  classical rotating matter respectively, computed in \cite{Barbieri:2005kp}. However, as we mentioned in previous section the scalar and spin dipole terms are  universal and follow from the soft behaviour of gravitational interactions. Therefore the wave cannot probe the nature of the compact object to this order in spin, which explains why the same result is obtained for a rotating star \cite{PhysRevD.16.237}. On the other hand, for higher multipoles our results show the genuine signature of the Kerr black hole in the exponentiation of $a\cdot q$, a remnant of the Newman-Janis shift \cite{Arkani-Hamed:2019ymq}. In \cref{sec:matching_BHPT_QFT} we explicitly show the matching of the scalar amplitude  \eqref{eq:scalar_kerr},  to the result from the  BHPT computation, up to order $(a\omega)^5$.

We can easily obtain the corresponding cross sections \eqref{eq:classical cross section} as before. For simplicity let us focus on the \textit{polar scattering}, where the momentum of the incoming wave is aligned with the spin direction. \footnote{For general spin orientation, the cross section can be easily computed using the amplitudes \eqref{eq:scalar_kerr}, \eqref{photon-kerr} and \eqref{eq:pppht}, for the different helicity scenarios. }, where the spin 3-vector is aligned with the impinging wave, e.g. $a_x=a_y=0$.  In this scenario, the scattering enjoys cylindrical symmetry and the cross-section simplifies drastically at finite scattering angle. We obtain

\begin{equation}\label{eq:polarcrallh}
    \frac{d\sigma}{d\Omega}= G^2 M^2 \frac{\cos^{4h}(\theta/2)}{\sin^4(\theta/2)}\left[\cosh{\left(2a_z\omega\sin^2(\theta/2)\right)}-h \sinh{\left(2a_z\omega\sin^2(\theta/2)\right)}\right]^2\,,h\leq 1\,.
\end{equation}
Notice that for $a_z=0 $, we recover the amplitudes for the scattering of waves off the SBH \eqref{eq:crforh}.

Let us now study the behaviour of our classical amplitudes in the small scattering angle, e.g. $\theta\to 0$ regime, which we refer to as eikonal limit. It can easily be shown that in this limit, for any fixed incoming helicity $h$ the scattering amplitude behave as

\begin{equation}\label{eq:eikoApp}
\boxed{    \langle A_{4,++}^{S\to \infty,h}\rangle=\frac{M^2\kappa^2}{\theta^2}\left[\cosh(a_x\omega\theta)-\frac{ia_y}{a_x} \sinh(a_x\omega\theta) +\ldots\right]}
\end{equation}
where $\ldots$ stands for terms  of the form $\omega^n \theta^m$ with $m>n$, which are non-universal and generically depend on $h$. Moreover, as in all cases we found $ \langle A_{4,--}\rangle = \left[\langle A_{4,++} \rangle^*\right]_{\omega\rightarrow-\omega}$, we conclude that the eikonal limit is universal. Thus, this extends the Newtonian universality found in \eqref{eq:newtonian} to all orders in the spin for Kerr background. We will show that this behaviour is indeed controlled by null geodesics in the Kerr background in  \cref{Eikonal-app}.

As in the previous section, the soft theorem of gravitons also guarantees that the off-diagonal elements of the scattering matrix \eqref{eq:scattering-matrix-helicities} vanish. This means that helicity is conserved in the scattering process. However, the fact that the diagonal elements $\langle A_{++}\rangle$,$\langle A_{--}\rangle$ are not equal induces a `shear' in the $2\times 2$ scattering matrix. We now  characterize this phenomena which emerges due to spin: The spinning motion of the Kerr black hole induces a partial  polarization of the waves.

\subsection{Spin-induced Polarization}\label{sec:polapp}

As anticipated, the induced polarization after the scattering will  be reflected in the difference between elements of the scattering matrix \eqref{eq:scattering-matrix-helicities}, which in turn  shows how waves of opposite circular polarization, scatter different from the Kerr BH. To see this explicitly we compare the scattering cross-sections for a left ($+$) and right ($-$) circularly polarized incoming wave:

\begin{equation}\label{eq:crsepm}
    \begin{split}
         64\pi^2 M^2 \frac{d\langle \sigma_+\rangle}{d\Omega}&=\langle A_{++} \rangle \langle A_{++} \rangle^* + \langle A_{+-} \rangle \langle A_{+-} \rangle ^*  \\
   64\pi^2 M^2 \frac{d\langle \sigma_-\rangle}{d\Omega}&=\langle A_{--} \rangle \langle A_{--} \rangle^* + \langle A_{-+} \rangle \langle A_{-+} \rangle ^* 
    \end{split}
\end{equation}
We have found in the previous examples that opposite helicity amplitudes are related via $ \langle A\rangle \to \langle A \rangle^*$, accompanied by the expected time reversal $\omega\to -\omega$, map. This is more transparent in the  spinor-helicity formalism, and can be seen as a consequence of CPT/crossing symmetry: Opposite helicities are related by chiral (i.e. complex) conjugation in the amplitude \cite{BCGV}. This induces a parity transformation which flips the sign of $a^\mu$, which corresponds to a pseudovector as it describes the orientation of the rotating black hole. Due to the fact that spin only enters through the combination $a\omega$ the map $a^\mu \to -a^\mu$ is of course equivalent to $\omega\to -\omega$.

From the above discussion, using \eqref{eq:crsepm}, we easily conclude that

\begin{equation}\label{eq:symcr}
    \frac{d\langle \sigma_+\rangle}{d\Omega}=   \left[ \frac{d\langle \sigma_-\rangle}{d\Omega}\right]_{(\omega\to-\omega)}
\end{equation}
Following \cite{PhysRevD.16.237,Dolan:2008kf,Barbieri:2005kp} we also introduce the polarization measurement

\begin{equation}\label{eq:polarization-formula}
    \mathcal{P}=\frac{\frac{d\langle\sigma_+\rangle}{d\Omega}-\frac{d\langle\sigma_-\rangle}{d\Omega}}{\frac{d\langle\sigma_+\rangle}{d\Omega}+\frac{d\langle\sigma_-\rangle}{d\Omega}}\,.
\end{equation}
According to \eqref{eq:symcr} the numerator of the polarization depends only on odd powers of $a\omega$ in the cross-section, and in particular vanishes for the Schwarzschild case. For the \textit{polar scattering} case considered in \eqref{eq:polarcrallh} we find

\begin{equation}\label{eq:polariation}
    \mathcal{P}^{(h)} =  -h \sinh{\left(4a_z|\omega|\sin^2(\theta/2)\right)} \left[\cosh^2{\left(2a_z|\omega|\sin^2(\theta/2)\right)} +h^2\sinh^2{\left(2a_z|\omega|\sin^2(\theta/2)\right)}\right]^{-1},
\end{equation}
which for $\theta\rightarrow 0$ becomes $ \mathcal{P}^{(h)} =  -ha_z|\omega|\theta^2$ and thus recovers the results in \cite{PhysRevD.16.237,Barbieri:2005kp}. Naively we might expect that equation \eqref{eq:polariation} also holds true for $h=2$, as was miss-concluded in \cite{Guadagnini:2008ha,Barbieri:2005kp} to linear order in  spin. This would be the case if to compute the scattering amplitude, we consider only the graviton exchange diagram between the Kerr background and the gravitational wave. In such a case however, the resulting amplitude fails to be gauge invariant, and therefore the additional diagrams need to be  consider. We expand on this issue in Part II. 

A non-trivial polarization implies that different helicity waves scatter by different angles. For instance, linearly polarized incoming waves become elliptically polarized after the scattering process \cite{PhysRevD.16.237}. We can obtain the scattering angles by integrating the cross section as in \eqref{eq:newtonian}: Writing $d\Omega=\sin\theta d\theta d\phi $ we have

\begin{equation}
    \sin(\theta)\frac{d\langle \sigma_{\pm} \rangle}{d\Omega} = b_\pm (\theta,\phi) \frac{\partial b_{\pm}}{\partial \theta}(\theta,\phi)\Longrightarrow b^2_{\pm}(\theta,\phi) = 2\int^\pi_\theta \frac{d\langle \sigma_{\pm} \rangle}{d\Omega}\sin(\theta')d\theta'\,.
\end{equation}
The angular splitting $\delta$ between both helicities is then computed by solving the condition
\begin{equation}\label{eq:angular-splitting}
    b^2_+(\theta,\phi)=b^2_{-}(\theta+\delta,\phi) = b^2,
\end{equation} 
for $\delta\ll \theta \approx \frac{4GM}{b}\ll 1$ (we also recall that our coordinate system is chosen so that $\phi=0$). To linear order in the spin of the black hole we find
\begin{equation}\label{eq:splb}
    \delta^{(h)} = 2\left(\frac{4GM}{b}\right)^3\left[a_z\omega h\log\left(\frac{b}{2GM}\right)+\frac{\omega}{2}\left(\frac{\pi  }{4}a_x  -h a_z  \right)\left( \gamma_E+\psi^{(0)}(1+2h)  \right) \right],\,\,\,h\leq 1\,.
\end{equation}
where $\psi^{(0)}(z)=\frac{d}{dz}\log(\Gamma(z))$ is the polygamma function. The result recovers that of \cite{PhysRevD.16.237} for $h=1$ and extends it to all helicities. In \cref{Eikonal-app} we will study the classical scattering angle $\theta$ in the eikonal regime, as controlled by geodesics of point-particles in Kerr. It will then be evident that the splitting \eqref{eq:splb} is a correction of order $\epsilon=GM\omega$ to the \textit{subleading} eikonal angle $\sim G^2M^2a/b^3$. This reflects that the angular splitting does not admit a point-particle interpretation, indeed the corrections are non-universal and depend on the helicity of the wave. If one insists in such a particle-like description one should include the so-called Papatetrou-type forces \cite{10.2307/98893}.

\section{Classical plane wave scattering from BHPT}\label{teukolsky}

By implementing our classical limit, we have written down the generic form of the 4-point scattering amplitudes to all orders in spin and helicities $h=0,1/2,1$.  Our aim in this section is to match the associated Newman-Penrose scalars to the classical solutions of the wave equation in BH backgrounds. We will focus on the $h=0$ case to show how the QFT amplitudes can be matched perfectly. In Part II of this work we will obtain similar results for the case $h=2$.

We use black hole perturbation theory (BHPT) to compute the \textit{classical} scattering amplitudes and the differential cross section for a plane wave scattering off both the Schwarzschild and the Kerr black hole, considering the weak-coupling regime $GM\omega\ll  1$ (but not necessarily only the leading order). As mentioned, in this limit the black hole size is negligible with respect to the scattering wavelength and we expect the singularity to behave as point-like compact object. Further considerations required to treat the spinning singularity are given in the next sections.

\subsection{Scalar Waves on the Schwarzschild Black hole} \label{sec:toyexample}
We begin by reviewing the case of scattering of scalar waves in a Schwarzschild spacetime. This will illustrate the partial wave approach without the encumbering technical details of Kerr spacetime. %the full Teukolsky gravitational calculation. 
For more details on the Schwarzschild problem see \cite{Glampedakis:2001cx}.

The dynamics of a scalar field $\psi$ is governed by the scalar wave equation:
\begin{align}\label{eq:kgschw}
    \nabla^{\mu}\nabla_{\mu} \psi=0.
\end{align}
In Schwarzschild spacetime this can be reduced to a radial ODE with the separability ansatz
\begin{align}
    \psi(t,r,\vartheta,\varphi)=\sum_{l=0}^\infty\sum_{m=-l}^{l}e^{-i\omega t}Y_{lm}(\vartheta,\varphi)R_{lm}(r)
\end{align}
where $Y_{lm}(\vartheta,\varphi)$ are the spherical harmonics, and the radial mode functions $R_{lm}(r)$ are decoupled and satisfy the $h=0$,  Regge-Wheeler equation \eqref{eq:RWeq}
\begin{align}
    \frac{d^2 R_{lm}}{d r_*^2}+(\omega^2-V_l(r))R_{lm}=0.
\end{align}
Here $r_*= r+2 GM \log(\frac{r}{2GM}-1)$ is the tortoise coordinate and the effective potential is $V_l(r)=\frac{r-2GM}{r}\left(\frac{l(l+1)}{r^2}+\frac{2GM}{r^3}\right)$.
We seek a vacuum solution $\psi$ to the scalar wave equation consisting of an incoming plane wave $\psi^{\rm PW}$ and an outgoing scattered wave $\psi^{\rm S}$:  $\psi=\psi^{\rm PW}+\psi^{\rm S}$. We choose the plane wave to move along the $z$-axis of the spacetime so that asymptotically it takes the form 
\begin{align}
\psi^{\rm PW}\sim e^{-i \omega(t-z)}. \label{Eq:PWonaxis}
\end{align}
Crucially, in this expression $z=r_*\cos\vartheta$, which corresponds to the expression for a plane wave in flat spacetime with the replacement $r\rightarrow r_*$, accounting for the long range gravitational potential from the black hole \cite{Matzner1968}. The scattered wave will asymptotically be a purely outgoing wave, which written on a basis of spherical harmonics takes the form
\begin{align}
    \psi^{\rm S}\sim r^{-1}\sum_{lm}A_{lm}^{\rm S} Y_{lm}(\vartheta,\varphi)e^{-i\omega(t-r_*)} \,.\label{eq:asymtpwa}
\end{align}
The key observable from the scattering processes is the differential cross section, which measures the angular profile of the flux from the scattered wave. For scalar waves it is given by \cite{futterman88}
\begin{align}
    \frac{d\sigma}{d\Omega}&=\lim_{r\rightarrow\infty}r^2|\psi^S|^2\\
    &=|\sum_{lm}A_{lm}^{\rm S} Y_{lm}(\vartheta,\varphi)|^2\equiv |f(\vartheta,\varphi)|^2
\end{align}
Determining the amplitude function $f(\vartheta,\varphi)$ and the differential cross section requires determining $A^{\rm S}_{lm}$, the asymptotic amplitude modes of the scattered wave, which can be fixed by a given set of boundary conditions. For that we first have to put in the same footing $\psi^{\text{PW}}$ and $\psi^{\rm{S}}$; we proceed as follows: By projecting onto spherical harmonics, the plane wave \eqref{Eq:PWonaxis} can be written  as
\begin{align}
    \psi^{\rm PW}\sim r^{-1}\sum_{l=0}^\infty e^{-i\omega t}Y_{l0}(\vartheta,\varphi)\left(A^-_{l}e^{-i\omega r_*}+A^+_{l}e^{i\omega r_*}\right),
\end{align}
where $A^-_{l}=2\pi i \omega^{-1} (-1)^l Y_{l0}(0,0)$ and $A^+_{l}=-2\pi i \omega^{-1} Y_{l0}(0,0)$. Since the plane wave only has modes with $m=0$ the scattered wave will also only be nonzero for $m=0$, i.e. $A^{\rm S}_{l,m\neq0}=0$ and we will write $A^{\rm S}_{l}\equiv A^{\rm S}_{l0}$.
The total solution can then be written
\begin{align}
    \psi&=\psi^{\rm PW}+\psi^{\rm S} \\
    &\sim r^{-1}\sum_{l=0}^\infty e^{-i\omega t}Y_{l0}(\vartheta,\varphi)\left(A^-_{l}e^{-i\omega r_*}+(A^+_{l}+A^{\rm S}_{l})e^{i\omega r_*}\right) \\
    &=r^{-1}\sum_{l=0}^\infty e^{-i\omega t}Y_{l0}(\vartheta,\varphi)\left(A^-_{l}e^{-i\omega r_*}+A^+_{l}e^{2 i \delta_l}e^{i\omega r_*}\right).\label{Eq:PWpSW}
\end{align}
In the last line we have introduced the phase shift function $\delta_l$, which is nothing but the particular form of the S-matrix in this basis
\begin{equation}
    e^{2 i \delta_l}=1+A^{\rm S}_{l}/A^+_{l}
\end{equation}
 Note in particular that the imaginary part of $\delta_l$, for which the scattered wave will have dissipation in \eqref{Eq:PWpSW}, is related to absorption into the black hole \cite{Sanchez:1976xm}, and that if $\delta_l$ were purely real, there would be no absorption and the scattering would be purely elastic.

The total solution must be a homogeneous solution which has the physical boundary condition of being a purely `ingoing' at the horizon. This condition fixes the total solution up to a normalization constant (see Appendix~\ref{Teuk-App}), and in particular at radial infinity it can be written as
\begin{align}
    \psi\sim r^{-1}\sum_{l=0}^\infty Z_{l}e^{-i\omega t}Y_{l0}(\vartheta,\varphi)\left(B^{\rm inc}_{l0}e^{-i\omega r_*}+B^{\rm ref}_{l0}e^{i\omega r_*}\right) \label{Eq:psigeneralschw}
\end{align}
where $B^{\rm inc/ref}_{lm}$ are the incidence and reflection coefficients which can be calculated with standard techniques, see \eqref{Eq:Binc}-\eqref{Eq:Bref} for $a=0$ and $h=0$, i.e. specialize to scattering of scalar waves off the Schwarzschild BH. Here $Z_{l}$ is the normalization which we will determine presently. Equating \eqref{Eq:PWpSW} and \eqref{Eq:psigeneralschw} gives
\begin{align}
    Z_{l}&=\frac{A^-_{l}}{B^{\rm inc}_{l0}}, \\
    A^{\rm S}_{l}&=\frac{A^-_{l}B^{\rm ref}_{l0}}{B^{\rm inc}_{l0}}-A^+_{l} \\
    &=\frac{2\pi}{i\omega} Y_{l0}(0,0)\left(e^{2i\delta_l}-1\right),
\end{align}
 where $e^{2i\delta_l}=(-1)^{l+1}\frac{B^{\rm ref}_{l0}}{B^{\rm inc}_{l0}}$, and the amplitude function is 
\begin{equation}
    f(\vartheta,\varphi)=\frac{2\pi}{i\omega}\sum_{l=0}^{\infty} Y_{l0}(0,0)Y_{l0}(\vartheta,\varphi)\left(e^{2i\delta_l}-1\right).\label{ec:spinlesspartialwaves}
\end{equation}
This will be the fundamental equation to match in the case of the Schwarzschild black hole. Next we  will show how to compute this infinite sum in the low energy limit.

\subsubsection{Long wavelength scattering}

We now impose that the wavelength of the scattered wave is large relative to the natural length scales of the problem, namely $\lambda\gg GM$, or in terms of the frequency $\omega\ll\frac{1}{GM}$. In this limit standard techniques (see e.g. \cite{Sasaki:2003xr}) lead to series expansions for the phase function $e^{2i\delta_{lm}}$ in the small parameter $\epsilon=2G M \omega$. As will be appreciated in the following, this is not the Born (tree-level) approximation in QFT but also encompasses certain loop effects. 

As $\epsilon\to 0$, the phase function can be written in a well known form, see for instance \cite{Andersson:2000tf},
\begin{align}
    e^{2i\delta_{lm}}\sim e^{i\Phi}\frac{\Gamma(l+1-i\epsilon)}{\Gamma(l+1+i\epsilon)}+\mathcal{O}(\epsilon^2)
\end{align}
where $\Phi=2\epsilon\log(2\epsilon)-\epsilon\to 0$ is an overall irrelevant phase (independent of $l$). The contribution is the `Newtonian' piece in the sense that it dominates in the limit $\theta \to 0$, containing the expected divergence as we now show.

We need then to compute the spherically symmetric sum
\begin{align}
    f^{N}(\vartheta)=\frac{2\pi}{i\omega}\sum_{l=0}^{\infty}\frac{2l+1}{4\pi}\frac{\Gamma(l+1-i\epsilon)}{\Gamma(l+1+i\epsilon)}P_{l}(\cos\vartheta),
\end{align}
where the $m=0$ spherical harmonic reduced to the Legendre function $P_l(x)$. Note that only the phase term contributes in \eqref{ec:spinlesspartialwaves}. This sum has closed form
\begin{align}\label{eq:scalar_amplitude_BHPT-Schw}
    f^{N}(\vartheta)=GM\frac{\Gamma(1-i\epsilon)}{\Gamma(1+i\epsilon)}\sin\left(\frac\vartheta2\right)^{-2+i2\epsilon}.
\end{align}
As anticipated, this agrees with the Born approximation \eqref{eq:scalara40}, up to a phase depending on $\epsilon$ (which expectedly vanishes as $\epsilon\to0$). The factor

\begin{equation}
    \frac{\Gamma(1-i\epsilon)}{\Gamma(1+i\epsilon)}\,,
\end{equation}
is termed `Newtonian phase' in analogy with its electromagnetic counterpart. Crucially, it is not a phase for $\omega\in \mathbb{C}$. Its poles located at $\epsilon:=2GM\omega=i n\,,n\in \mathbb{N}$ are nothing but the spectrum of bounded states of the Newtonian problem. We will see how this phase can be recovered from tree-level amplitudes in \cref{sec:pwtoeikonal}.

\subsection{Scalar Waves on the Kerr Black Hole}
\label{sec:KerrPWs}
We are now ready to tackle the problem of scalar waves in Kerr via partial waves methods. Here the fundamental fact is the separability of the scalar wave equation into the so-called Teukolsky equations and its solution based on BHPT as we review in Appendix \ref{Teuk-App}.

 In order to obtain analytic solutions we will need the following considerations. In the spinning case there are two  dimensionless quantities which appear in our analysis: $G M \omega$ and $a/G M$. 
To reproduce the BH computation, we will consider the BHPT solution fixed by boundary conditions at the horizon, which means in principle we include non-linear terms in $GM\omega$ in our solution while keeping all orders in $a/GM$. Only after that do we take the linear order in $G$ contribution in order to match the tree-level amplitudes of \cref{sec:Section3}. Higher order terms in $G$ can also be matched to loop amplitudes (in the eikonal limit) as we illustrate in the next section.
 The expansion we shall adopt  then corresponds to a point-particle setup in which only the intrinsic structure of the BH, i.e. its spin multipoles, is preserved.  For this we consider the black hole characteristic size to be small as compared to both the wavelength of the waves $1/\omega$ and the `spin' radius $a$. That is to say, we take small $GM\omega$ and large $a/GM$. Of course, the latter of these requirements naturally reveals the ring-singularity as its radius $a$ grows bigger than the horizon itself, leading to a superextremal Kerr metric. However,  we expect these effects to be unobservable from large distances as the singularity becomes essentially point-like, so we will further consider $\omega \sim 1/r \ll 1/a$. Thus, in synthesis, we assume

\begin{equation}\label{eq:analytic_continuation}
    \omega\ll\frac{1}{a}\ll\frac{1}{GM},\qquad \lambda\gg a\gg GM.
\end{equation}
In practice we will determine the differential cross section as an expansion in small frequency at fixed $a/GM < 1$, where BHPT is applicable. We will then analytically continue all functions to $a/GM>1$, with the prescription described above. 

Throughout this section we will use $(t,r,\vartheta,\varphi)$ to denote standard Boyer-Lindquist coordinates. The main alteration to the previous discussion is that the separability ansatz solving \eqref{Eq:TeukMaster}, now relies on the spheroidal harmonics. That is,
\begin{align}
    \psi(t,r,\vartheta,\varphi)=\sum_{l=0}^\infty\sum_{m=-l}^{l}e^{-i\omega t}S_{lm}(\vartheta,\varphi;a\omega)R_{lm}(r)
\end{align}
where the radial functions satisfy the $h=0$ radial Teukolsky equation \eqref{Eq:RadTeuk}, whereas the  spheroidal harmonics are the solution to the angular equation \eqref{Eq:Slmeq}. They  admit a low frequency expansion whose leading order is simply the spherical harmonics
\begin{align}
    S_{lm}(\vartheta,\varphi;a\omega)\sim Y_{lm}(\vartheta,\varphi)+\mathcal{O}(a^2\omega^2).
\end{align}
but whose subleading orders contain, as expected, corrections related to the internal structure of the spinning body.

Let us first comment on the \textit{polar scattering } scenario. In this case the plane wave moves up the BH axis of rotation, which is aligned with z-axis of the coordinate system. All other details carry through as in Schwarzschild scattering and the amplitude function is now
\begin{align}\label{ec:polarspheroidal}
    f(\vartheta,\varphi)=\frac{2\pi}{i\omega}\sum_{l=0}^{\infty} S_{l0}(0,0;a\omega)S_{l0}(\vartheta,\varphi;a\omega)\left(e^{2i\delta_l}-1\right),
\end{align}
with $e^{2i\delta_l}=(-1)^{l+1}\frac{B^{\rm ref}_{l0}}{B^{\rm inc}_{l0}}$, where all dependence on the BH spin is now included in the corresponding coefficients $B^{\rm ref/inc}_{lm}$, given by \eqref{Eq:Binc}-\eqref{Eq:Bref}.

On the other hand, when the plane wave is incident along a vector in the $x$-$z$ plane at an angle $\gamma$ from  the $z$-axis (See \cref{fig:Initial}), the calculation follows the same procedure, with some more technical complications (see Appendix A of \cite{Glampedakis:2001cx}). The end result is that the amplitude function is
\begin{align}
    f(\vartheta,\varphi)=\frac{2\pi}{i\omega}\sum_{l=0}^{\infty} \sum_{m=-\infty}^{\infty}S_{lm}(\gamma,0;a\omega)S_{lm}(\vartheta,\varphi;a\omega)\left(e^{2i\delta_{lm}}-1\right), \label{Eq:fKerrGeneric}
\end{align}
with $e^{2i\delta_{lm}}=(-1)^{l+1}\frac{B^{\rm ref}_{lm}}{B^{\rm inc}_{lm}}$. Thus the phases acquire and additional dependence on the azimuthal quantum number $m$ conjugate to the angle $\varphi$. This will play an important role in the eikonal description of scattering, where it will be identified with the angular momentum component $L_z$.

\begin{figure}
     \centering
     \begin{subfigure}[b]{0.495\textwidth}
         \centering
         \includegraphics[width=\textwidth]{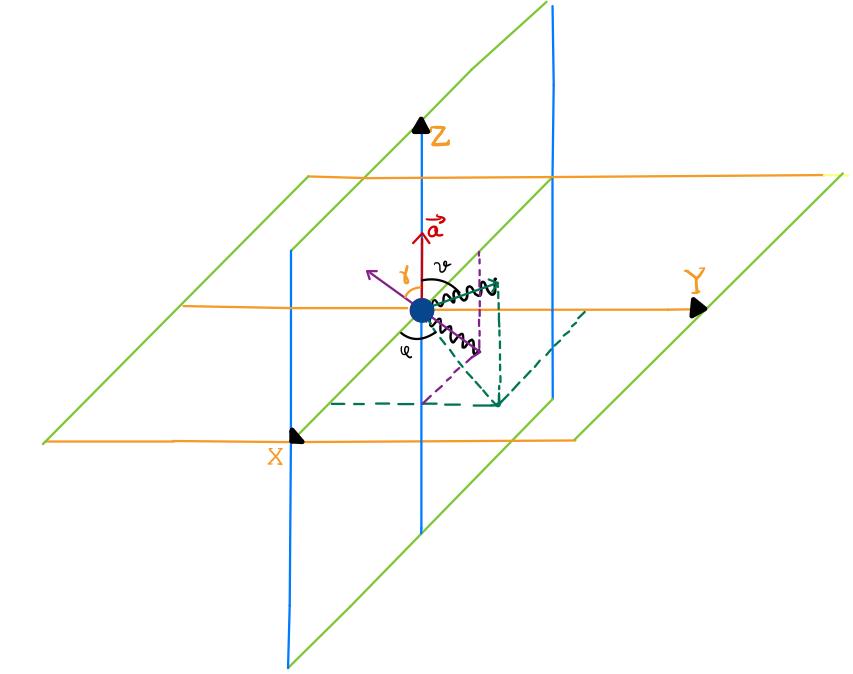}
         \caption{Initial BHPT setup}
         \label{fig:Initial}
     \end{subfigure}
     \hfill
     \begin{subfigure}[b]{0.495\textwidth}
         \centering
         \includegraphics[width=\textwidth]{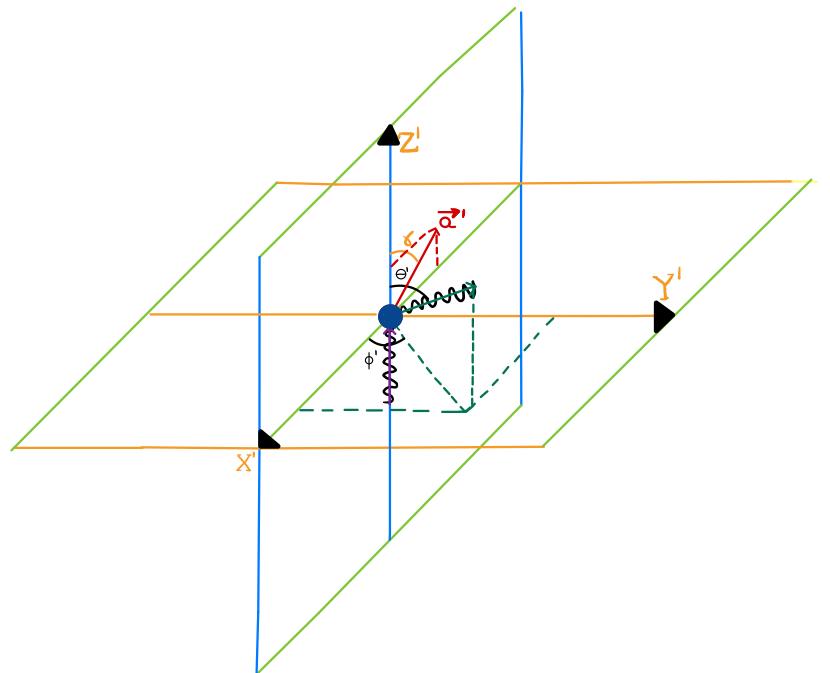}
         \caption{Rotated BHPT setup}
         \label{fig:rotated}
     \end{subfigure}
     \caption{Wave scattering in the  BHPT setup. (a) An incoming PW (purple) impinges on the BH at an angle $\gamma$ with respect to the direction of the BH spin ($\vec{a} = a \hat{z}$). The outgoing scatter wave (green) moves in a general direction with angles $\vartheta$ and $\varphi$ with respect to the $Z$ and $X$ axis respectively. (b) Rotated frame. In this frame, the incoming PW (purple) moves along the vertical axis, whereas the spin of the BH is rotated at an angle $\gamma$ with respect to $Z'$. The outgoing scatter wave (green) now moves in the general direction $\theta',\phi'$ with respect to the $Z'$ and $X'$ axis respectively. }
    %  \hfill
    %  \begin{subfigure}[b]{0.3\textwidth}
    %      \centering
    %      \includegraphics[width=\textwidth]{Figures/amplframe.jpg}
    %      \caption{$y=5/x$}
    %      \label{fig:five over x}
    %  \end{subfigure}
    %     \caption{Three simple graphs}
    %     \label{fig:three graphs}
    \label{fig:frames}
\end{figure}

\subsubsection{Higher order scattering}\label{sec:Hos}

Writing the phase function in the factorised form
\begin{align}
    e^{2i\delta_{lm}}= e^{i\Phi}\frac{\Gamma(l+1-i\epsilon)}{\Gamma(l+1+i\epsilon)}\beta_{lm},\label{Eq:phasetobeta}
\end{align}
higher order corrections in the spin of the BH  are captured within the function $\beta_{lm}$. These take the form
\begin{align}
    \beta_{lm}=1+\beta_{lm}^{(2)}\epsilon^2+\beta_{lm}^{(3)}\epsilon^3+\beta_{lm}^{(4)}\epsilon^4+\ldots
\end{align}

Here we have also included higher orders in $\epsilon$, as anticipated  above. The reason is the following: for generic modes the $\beta_{lm}^{(i)}$ are polynomials of $a^\star =a/GM$, which then makes the combinations $z^n=(\epsilon a^\star)^n =(2a\omega)^n$, of order $\mathcal{O}(G^0)$, indeed  contributing to the tree-level amplitude. We find this observation crucial to correctly match Teukonsly solutions to the spinning amplitudes of \eqref{sec:Section3}.

As mentioned above eq. \eqref{eq:analytic_continuation}, we will only be concerned with the $a^\star \gg1$ expansion of these functions, and so we give here only the leading order terms in the large-$a^\star$ expansion. For \textit{generic} values of $l$ (see below), they are given by
\begin{align}
    \beta_{lm}^{(2)}&\sim-\frac{im}{l(l+1)}a^\star+\mathcal{O}(1),\label{Eq:blm2}\\
    \beta_{lm}^{(3)}&\sim %\left(\frac{3 i m^2}{2 l (l+1) (2 l-1) (2 l+3)}-\frac{i}{2 (2 l-1) (2 l+3)}\right)q^2 
    -\frac{i  \left(l(l+1)-3 m^2\right)}{2 l (l+1) (2 l-1) (2 l+3)}a^{\star 2}+\mathcal{O}(a^\star),\label{Eq:blm3}\\
    \beta_{lm}^{(4)}&\sim  -\frac{i m \left(3 l^2+3 l-1-5 m^2\right)}{6 (l-1) l (l+1) (l+2) (2 l-1) (2 l+3)}a^{\star 3}  +\mathcal{O}(a^{\star 2}),\label{Eq:blm4}\\
    \beta_{lm}^{(5)}&\sim \frac{i}{8 l (l+1) (2 l-3) (2 l-1)^3 (2 l+3)^3 (2 l+5)} \bigg(l (l+1) (12 l^4+24
   l^3-l^2-13 l+15\nonumber\\
   &-2 \left(100 l^4+200 l^3+13 l^2-87 l+45\right) m^2+9 \left(28 l^2+28 l-5\right) m^4)\bigg) a^{\star 4}  +\mathcal{O}(a^{\star 3}),\label{Eq:blm5}\\
   \beta_{lm}^{(6)}&\sim -\frac{i m }{40 (l-2) (l-1) l (l+1) (l+2)
   (l+3) (2 l-3) (2 l-1)^3 (2 l+3)^3 (2 l+5)}\nonumber \\
   &\bigg(140 l^8+560 l^7+35 l^6-1855 l^5-958 l^4+1829 l^3+364 l^2-953 l+138\nonumber \\
   &-10 \left(84 l^6+252 l^5-135 l^4-690 l^3-28 l^2+359 l-192\right) m^2\nonumber \\
   &+\left(828 l^4+1656 l^3-2009 l^2-2837 l-438\right)
   m^4\bigg) a^{\star 5}  +\mathcal{O}(a^{\star 4}),\label{Eq:blm6}
\end{align}
Since the $\beta_{lm}^{(i)}$ are simply polynomials in $a^{\star}$, these expressions are identical to the highest power of $a^{\star}$ in the full expressions. We highlight here the important feature that in this limit each $\beta_{lm}^{(i)}$ is purely imaginary, which results in the  phase function $\delta_{lm}$ being purely real. As mentioned, this corresponds to the absence of horizon absorption \cite{Sanchez:1976xm,Dolan:2008kf}. While we do not have a definition of horizon absorption in the limit  $a^\star \gg1$, we assume the analytic continuation from $a^\star <1$ preserves these features.

These results are valid for generic values of $l$, meaning that they are indeed incorrect for some $l<l_{\textrm{min}}$. In fact, it can be seen that they contain poles at such low values of  $l$, reminiscent of anomalous contributions in Regge theory \cite{Gribov:2003nw}. As we will 
see in \cref{Eikonal-app}, and as will be preponderant in Part II,  generic values of $l$ are associated to factorization diagrams, whereas certain small values of $l$ (or anomalous) shall be associated to contact terms in the QFT amplitude. We now discuss the anomalous contributions.

\subsubsection*{Anomalous Terms and Absorption}

The generic-$l$ expressions need to be supplemented by explicit computation of low-$l$ values which contain anomalous terms. In particular \eqref{Eq:blm2}-\eqref{Eq:blm6} are valid for all $l\geq3$. 
For $l=2$: $\beta_{lm}^{(i\geq5)}$ are incorrect, for $l=1$: $\beta_{lm}^{(i\geq3)}$ are incorrect and for $l=0$ none are correct. While the full expressions for the  anomalous values will be provided in the supplemental files of the \texttt{arXiv} version of this paper,
we will examine closely particular terms to illustrate the complications which arise. For example, consider the first two terms for $l=m=0$:
\begin{align}
    \beta_{00}^{(2)}&=-(1+\hat\kappa-\frac{11}{6}i\pi), \\
    \beta_{00}^{(3)}&=\frac{i}{18}(3 {a^\star}^2+(-33+18 i \pi ) \hat\kappa-36 \log(2\epsilon\hat\kappa)+11 \pi ^2+18 i \pi -36 \gamma_\mathrm{E} +51),
\end{align}
where $\hat\kappa=\sqrt{1-a^{\star 2}}$ and $\gamma_\mathrm{E}$ is Euler's constant.
We  see immediately the dependence on $a^\star$ is no longer purely polynomial due to the appearance of $\hat\kappa$. For $\beta_{00}^{(2)}$ the leading  order contribution in the limit $a^\star \gg1$ comes from the $\hat\kappa$ term, and notably is \textit{real} when $a^\star <1$. Thus we interpret it as being an absorption effect. Anticipating the matching to the quantum calculation which manifestly has a conservation of incoming and outgoing radiation (the tree-level amplitudes are real away from singular kinematics) we will remove  the absorption. 

Our procedure to remove absorption is simply to take the real part of $\delta_{lm}$\footnote{In practice we will ignore the factor of $\frac{\Gamma(l+1-i\epsilon)}{\Gamma(l+1+i\epsilon)}$ in \eqref{Eq:phasetobeta} in this procedure since it does not affect the leading-order-in-G.}. This is expected to be equivalent to setting reflective boundary conditions on the horizon but we have not explicitly checked this. Once absorption is removed in this manner, we can analytically continue the result to the $a^\star \gg1$ limit. Defining the new `conservative' scattering by $\tilde{\beta}_{lm}$, our procedure has the  desired effect on the examples given above, removing the dependence on $\hat\kappa$, and moreover having the leading order  term in the $a^\star \gg1$ limit come purely from polynomial dependence in $a^\star$:
\begin{align}
   \tilde{\beta}_{00}^{(2)}&= \frac{11}{6}i\pi, \\
   \tilde{\beta}_{00}^{(3)}&=\frac{1}{18} i \left(3 a^{\star 2}-33 \hat\kappa-36 \log (2 \epsilon\hat\kappa )+11 \pi ^2-36 \gamma_\mathrm{E} +51\right).
\end{align}
When $m\neq 0$, more complicated functions arise changing this simple behavior. For example, 
\begin{align}\label{eq:b11_appearence_polygamma}
    \beta_{11}^{(4)}&=-\frac{1}{360} i \left(2 {a^\star}^3-65 i {a^\star}^2+160 {a^\star}-10 i\right)+\frac{ 12870 i a^{\star 2}+65100 {a^\star}+78037 i}{189000}\pi+\frac{i  {a^\star}}{9} \gamma_\mathrm{E} \nonumber\\
    &+\frac{1}{18} i {a^\star} \log (2 \epsilon \hat\kappa)- \left(\frac{361}{1800}+\frac{7 i {a^\star}}{90}\right)\pi ^2+\frac{1 }{36}\hat\kappa+\frac{1}{18} i a^\star \psi ^{(0)}\left(i {a^\star} /\hat\kappa\right).
\end{align}
where $\psi ^{(0)}(z)$ is the digamma function. We have not found a simple prescription to take the large $a^\star$ limit of the polygamma function: Since $\psi^{(n)}(z)$ diverges at $z=-1$ but is finite at $z=1$, the polygamma function can contribute or not to the tree-level amplitude in this  limit depending on the branch that we choose for $\hat{\kappa}$, as the argument becomes $i{a^\star}/\hat{\kappa}\to \pm 1$. Explicitly, the two branches give
\begin{align}
 &  \frac{1}{36} i {a^\star} \psi ^{(0)}\left(i {a^\star} /\hat\kappa_+\right)\sim-\frac{1}{36} i \gamma  a^\star+\mathcal{O}({a^\star}^{-1}), \\
    &\frac{1}{36} i {a^\star} \psi ^{(0)}\left(i {a^\star} /\hat\kappa_-\right)\sim\frac{i {a^\star}^3}{18}+\mathcal{O}({a^\star}),
\end{align}
Notably, giving a real contribution to the phase, the second branch can explicitly contribute to the ${a^\star}^3$  order term of the tree-level scattering amplitude (recall $\beta_{lm}$ scales like $\mathcal{G^0}$). To account for this ambiguity let us write\footnote{Of course, according to our prescription we should take the real part $\psi ^{(0)}\left(i {a^\star} /\hat\kappa\right)+\psi ^{(0)}\left(-i {a^\star} /\hat\kappa\right)$ before the large ${a^\star}$ limit, which can certainly account for the contribution of both branches. However, even this combination does not contribute to the limit  $|a^\star|\to \infty$ in a generic (complex) direction, as opposed to the polynomial expansions we have obtained previously, so it still appears natural to remove it.}

\begin{align}\label{eq:b11_appearence_polygamma}
    \beta_{11}^{(4)}&=\beta_{11,\textrm{reg}}^{(4)}+\frac{1}{18} i {a^\star} \psi ^{(0)}\left(i {a^\star} /\hat\kappa\right).
\end{align}
 We find similar behaviour with the  polygamma functions for $m=-1$, as well as at higher powers of $\epsilon$ for different $l$-values, where we define the regularized values $\beta_{lm,\textrm{reg}}$ by simply excluding the contributions from the polygamma functions $\psi^{(n)}$. As it turns out, we show in the next section how the contribution from the regularized values $\beta_{lm,\textrm{reg}}$, or more precisely their real parts $\tilde{\beta}_{lm,\textrm{reg}}$, provides a perfect match to the Born amplitudes obtained in \cref{sec:Section3}. On the other hand, the contributions from  $\psi^{(n)}$ lead to additional contact terms, hence can also be added to the Born amplitudes without spoiling their factorization properties, as we will see bellow.

\subsection{Matching procedure: BHPT and QFT amplitudes }
\label{sec:matching_BHPT_QFT}

In order to match our QFT amplitudes with the results from BHPT, it is convenient to now project the previous amplitude function \eqref{Eq:fKerrGeneric} onto \textit{spherical} harmonics
\begin{equation}
f(\vartheta,\varphi)=\sum_{lm}Y_{lm}(\vartheta,\varphi)f_{lm}(\gamma),
\label{Eq:fSpher}
\end{equation}
where $f_{lm}=f^N+f^{(1)}_{lm}z+f^{(2)}_{lm}z^2+\ldots$ and $z=a^{\star}\epsilon={\textcolor{black}{2}}a\omega$. While there are some subtleties in the projection as discussed in Sec. 4.3.1. of \cite{Dolan:2008kf}, we will omit such details here.

An important feature emerges for \textit{polar scattering}, as described around \eqref{ec:polarspheroidal}, which is obtained by setting $\gamma=0$. One finds that only the modes $f^{(i)}_{l0}$ are non-trivial and

\begin{equation}
f^{(i)}_{l0}(0)=0\,,\,\,\textrm{ for  }\,\, i=2k+1\,\, \textrm{  or for  } i\leq l. 
\end{equation}
This means that, other than the Newtonian term, for  \textit{polar scattering} the infinite sum over the spherical harmonics reduces to a finite sum for each power of $z$. This has been noted in \cite{Dolan:2008kf} for the case of GW scattering in Kerr and will be proved in Part II of this work starting from the QFT framework.

In the off-axis case where $\gamma\neq0$, no such  simplification occurs, and we find it convenient to compare with the 4-pt amplitudes in \cref{sec:Section3}, more precisely the scalar amplitude \eqref{scalar-kerr}, by working mode-by-mode. To do this we first need to align our coordinates by a rotation (see \cref{fig:frames}).

Let us see how this works. The amplitude function \eqref{Eq:fSpher} is written as a sum over  spherical harmonics $Y_{lm}(\vartheta,\varphi)$,
where $(\vartheta,\varphi)$ is the direction of the outgoing wave in a coordinate system where the spin direction is $\vartheta=0$, i.e. the $+Z$ direction, and the incoming wave is in the direction $(\vartheta,\varphi)=(\gamma,0)$ (in the $X$-$Z$ plane), so that $\gamma$ is the angle between the spin and the incoming wave.

Now let us rotate our $(\vartheta,\varphi)$--$(X,Y,Z)$ coordinate system about the $Y$ axis (the same as the new $Y'$-axis) by an angle $\gamma$, to bring the incoming wave direction to the new $+Z'$ direction, and call the new coordinates  $(\theta',\phi')$--$(X',Y',Z')$.  This is still not the same as the $(\theta,\phi)$--$(x,y,z)$ coordinate system used in \cref{fig:ampl}, but now the $z$-axes are the same. 
The rotation of spherical harmonics is known to be accomplished by
\begin{equation}
Y_{lm}(\theta',\phi')=\sum_{m'}D^{l*}_{mm'}(\gamma)\;Y_{lm'}(\vartheta,\varphi)\,,
\end{equation}
where $D^{l*}_{mm'}$ is the (complex conjugate) Wigner $D$-matrix with Euler angles $(0,\gamma,0)$,
\begin{equation}
D^{l*}_{mm'}(\gamma)=D^{l*}_{mm'}(0,\gamma,0)=(-1)^{m'}\sqrt{\frac{4\pi}{2l+1}}{}_{-m'}Y_{lm}(\gamma,0).
\end{equation}
Now the amplitude \eqref{Eq:fSpher} takes the form
\begin{equation}\label{Flm}
f=\sum_{lm}Y_{lm}(\theta',\phi')f'_{lm}(\gamma),
\end{equation}
with
\begin{equation}
f'_{lm}(\gamma)=\sum_{m'}D^{l*}_{m'm}(\gamma)\; f_{lm'}(\gamma),
\end{equation}
where we relabeled $m\leftrightarrow m'$ after substituting.
Now, in this new $(\theta',\phi')$ coordinate system, with corresponding $(X',Y',Z')$, the spin vector is $\vec a=a(-\sin\gamma,0,\cos\gamma)$, the incoming wave vector is $\vec k_2=(0,0,\omega)$, and the outgoing wave vector is $\vec k_3=\omega(\sin\theta'\cos\phi',\sin\theta'\sin\phi',\cos\theta')$.

Finally, we have the third $(\theta,\phi)$--$(x,y,z)$ coordinate system used in \cref{sec:2}, where $\vec k_2$ is in the $z$ direction (same as $Z'$ direction) and $\vec k_3$ is in the $x$-$z$ plane at an angle $\theta$ from $\vec k_2$, and this is the same $\theta=\theta'$ from the second coordinates.  To translate the result \eqref{eq:scalar_kerr} into the $(\theta',\phi')$ coordinates, we use $\theta=\theta'$ and
\begin{equation}\label{eq:axayaz}
a_z=a\cos\gamma, \quad
 a_x=-a\sin\gamma\cos\phi', \quad
 a_y=- a\sin\gamma\sin\phi',
\end{equation}
as can be confirmed by comparing the values of $\vec a\cdot\vec k_2$, $\vec a\cdot\vec k_3$ and $\vec k_2\cdot\vec k_3$.
  
 It is most convenient to compare our amplitudes results from \cref{sec:Section3} using $(\theta,\phi)$ to our BHPT results from \cref{sec:Hos} using $(\vartheta,\varphi)$ by comparing the amplitudes of their spherical harmonic modes in the intermediate $(\theta',\phi')$ coordinates as in \eqref{Flm}.  
 
We recall that the amplitude function $f$ at the leading order in $\epsilon$ (at fixed $a\omega=a^\star\epsilon/2$), coming from the tree-level scattering amplitude \eqref{eq:scalar_kerr}, is
  \begin{equation}\label{eq:f0Kerr}
f=\frac{1 }{ 2\sin^2\left(\frac{\theta}{2}\right)}\left[ \cosh{a{\cdot}q}- i\omega a_y\sin\theta  \frac{\sinh{a{\cdot}q}}{a{\cdot}q} \right],       
\end{equation}
with $q{\cdot}a= -\omega( a_x\sin(\theta)-2 a_z\sin^2(\theta/2))$, which is expressed in terms of $(\theta',\phi')$ by using $\theta=\theta'$ and \eqref{eq:axayaz}.  Its mode amplitudes $f'_{lm}$ from \eqref{Flm} are given by integrals over the 2-sphere,
\begin{equation}
f'_{lm}(\gamma)=\int d\Omega'\,Y^*_{lm}(\theta',\phi') f(\gamma,\theta',\phi'),
\end{equation}
and depend only on the angle $\gamma$ between the incoming momentum and the spin, and on the parameters $\epsilon$ and $a^\star$.  In general, let us write these as an expansion in $\epsilon$, focusing on the leading order in the large $a^\star$ expansion at each order in $\epsilon$,
\begin{equation}
f'_{lm}=\sum_{n=0}^\infty \epsilon^n \Big[f'_{lm,n}a^\star{}^n+\mathcal O(a^\star{}^{(n-1)})\Big].
\end{equation}
We find that this pattern also holds for the analytically continued BHPT amplitudes in the large $a^\star$ expansion.  At linear order in spin, from \eqref{eq:f0Kerr}, we find
\begin{alignat}{3}
\begin{aligned}
f'_{00,1}&=0,
\\
\{f'_{1m,1}\}&=\frac{1}{2}\sqrt{\frac{3\pi}{2}}\sin\gamma\{1,0,1\},
\\
\{f'_{2m,1}\}&=\frac{1}{2}\sqrt{\frac{5\pi}{6}}\sin\gamma\{0,1,0,1,0\},
\\
\{f'_{3m,1}\}&=\frac{1}{4}\sqrt{\frac{7\pi}{3}}\sin\gamma\{0,0,1,0,1,0,0\},
\end{aligned}
\end{alignat}
and so on, with $m=\{-l,\ldots,l\}$.  From \eqref{eq:f0Kerr} at quadratic order in spin, we find
\begin{alignat}{3}
\begin{aligned}
f'_{00,2}&=\frac{\sqrt{\pi}}{16}(3+\cos 2\gamma),
\\
\{f'_{1m,2}\}&=\{\frac{1}{4}\sqrt{\frac{\pi}{6}}\sin 2\gamma,-\frac{1}{16}\sqrt{\frac{\pi}{3}}(1+3\cos 2\gamma),-\frac{1}{4}\sqrt{\frac{\pi}{6}}\sin 2\gamma\},
\\
\{f'_{2m,2}\}&=\frac{1}{16}\sqrt{\frac{5\pi}{6}}\sin^2\gamma\{1,0,0,0,1\},
\\
\{f'_{3m,2}\}&=\frac{1}{16}\sqrt{\frac{7\pi}{30}}\sin^2\gamma\{0,1,0,0,0,1,0\},
\\
\{f'_{4m,2}\}&=\frac{1}{16}\sqrt{\frac{\pi}{10}}\sin^2\gamma\{0,0,1,0,0,0,1,0,0\},
\end{aligned}
\end{alignat}
and so on.  Remarkably, also continuing to large values of $l$, we find that all of these $(\epsilon a^\star)^1$ and $(\epsilon a^\star)^2$ terms in the mode amplitudes from the (minimal) tree-level scattering amplitude \eqref{eq:f0Kerr} precisely match those computed from the analytically continued (and absorption-removed) BHPT theory amplitudes as described in \cref{sec:Hos}.

At the $(\epsilon a^\star)^3$ and $l\leq 2$ the BHPT computation leads to anomalous terms as explained around eq. \eqref{eq:b11_appearence_polygamma}. Taking $\beta_{lm}\to\beta_{lm,\textrm{reg}}$, that is, discarding the extra terms arising from polygamma functions we obtain a perfect match to the Born amplitudes. 

Since the Born amplitudes can be obtained directly by solving the classical Klein-Gordon equation in the limit $GM\omega\to 0$, it makes sense to regard our prescription as a recipe for obtaining the point-particle limit. However, it is also of interest to understand the contributions from the polygamma functions more in detail. As it turns out, the fact that the corrections appear only for $l\leq 2$ enables us to write down a deformation of the Born amplitude \eqref{eq:f0Kerr} that resums these terms

\begin{equation}
f_{\textrm{anomalous}}=\frac{1 }{ 2\sin^2\left(\frac{\theta}{2}\right)}\left[ \cosh{a{\cdot}q}- i\omega a_y\sin\theta
\left( \frac{\sinh{a{\cdot}q}}{a{\cdot}q}+\Delta f \right)+\mathcal O(a^7)\right].
\end{equation}
where the anomalous contributions can be found to be
\begin{equation}
    \Delta f= \frac{q^2a^2}{3} \left(1{-}\cot^2(\theta/2)\frac{q^2a^2}{15}{+}\frac{(a{\cdot}q)^2}{10}\right)\,\,\,,\qquad q^2=-4\omega^2\sin^2(\theta/2)\,. 
\end{equation}

As anticipated, the correction corresponds to a contact term, with no factorization residues in $q^2$. As we will see when discussing the eikonal approximation, this just reflects the fact that the anomalous contribution is not present for large $l$. On the other hand, contact terms can be added to the four-point amplitude without spoiling the three-point on-shell structures that have been previously matched to the Kerr metric. In fact, as anticipated below \eqref{eq:Fouh}, the first term $\sim a^3q^3$ can be easily incorporated in the effective vertex as a $\mathcal{O}(q^2)$ correction.

\section{A classical Wave-Particle duality and the eikonal in Kerr}\label{Eikonal-app}

Unlike the Schwarzschild case, the results \eqref{scalar-kerr}-\eqref{eq:pppht} for the scattering of waves of diverse helicities in the Kerr background do not share a universal pattern. Moreover, for scattering of gravitational waves, it will become clear in Part II that diverse effective operators emerge and lead to an even more exotic result for $h=2$.

However, we have found in equation \eqref{eq:eikoApp} that a universality class arises for the Kerr black hole in the forward limit $\theta\to 0$, which we refer to as the \textit{high-energy, eikonal} or \textit{geometrical optics} limit.\footnote{The eikonal limit has been studied with renewed interested for both massive and massless particles since it allows for a clean passage to obtain classical observables \cite{AccettulliHuber:2020oou, DiVecchia:2021bdo, DiVecchia:2020ymx,DiVecchia:2019myk,Chen:2021huj,Saotome:2012vy,DiVecchia:2019kta,Bern:2020gjj}. It is interesting however that in the case of wave scattering we are able to define a classical limit  for finite values of $\theta$. A complementary picture in this context of wave scattering is to be provided in \cite{CGKO}.} This holds for any helicity configuration, including gravitational waves as we will show. The reason is that, as we will see, such regime corresponds to a particle-like behaviour of the scattered waves and the phases computed in the previous section encode classical trajectories in the sense of the WKB/optical approximation. Therefore, the universality reflects that, classically, massless particles follow null geodesics irrespective of their internal structure. From the QFT perspective, the eikonal is controlled by factorization channels where gravitons couple to the massless particle in a universal way, reflecting the equivalence principle \cite{Guevara:2018wpp}.

In this section we aim to make the previous discussion precise. We  first show how the partial wave series reduces to the eikonal amplitude. The eikonal phase is obtained from the partial waves in the large $l$ limit, or equivalently `high-energies' $\omega b\to \infty$ for an impact parameter $b$. At leading order in $\epsilon=GM\omega \to 0$ the eikonal is linked directly to Born amplitude computed previously, but further provides information to all orders in $G$ such as the Newtonian phase. Moreover, the partial waves computed in the previous section also incorporate corrections in $\epsilon$: We match such corrections to the subleading eikonal phases computed recently in \cite{Guevara:2017csg,Liu:2021zxr,Kosmopoulos:2021zoq,Hinderer:2008dm} from loop amplitudes and higher orders in spin. This provides an important cross-check of different results.

As will be argued below, the eikonal phase, even at subleading (loop) orders in $\epsilon$, is determined solely from the radial action of null geodesics in Kerr in the spirit of the WKB approximation. Because of this, using Hamilton-Jacobi theory, it can be used to compute two important classical observables: the scattering angle and the Shapiro time delay, which reflects the underlying unitarity of the effective theory. We incorporate spin corrections to these observable in two regimes: Polar and equatorial orientations of the incoming wave.

\subsection{From Partial Waves to Eikonal}\label{sec:pwtoeikonal}

    The eikonal amplitude can be easily obtained from the $\theta \to 0$ limit of partial waves, and is dominated by the contributions at large angular orbital momentum $l$. Let us illustrate this with the spinless case $a=0$ for simplicity.\footnote{This closely follows more detailed discussions that can be found in e.g. \cite{FORD1959259,brittin1959lectures,Gribov:2003nw,BETHE1951,Glampedakis:2001cx,Andersson:2000tf,Jauch:1976ava}, where the argument holds at least to subleading order in $\theta\to 0$.} Consider the partial wave amplitude \eqref{ec:spinlesspartialwaves},

\begin{equation}\label{fthpl}
    f(\theta)=\frac{1}{2i\omega}\sum_{l=0}^\infty(2l+1)(e^{2i\delta_l}-1)P_l(\cos(\theta))\,,
\end{equation}
In the $\theta \to 0$ limit the scattered momenta is close to the axis of the incoming wave. Spherical waves are then approximated by cylindrical waves via

\begin{equation}
    P_l(\cos(\theta))= J_0 \left((2l+1)\sin(\theta/2)\right)+ \mathcal{O}(\sin^2(\theta/2) )
\end{equation}
Such relation is most accurate at large $l$ which will provide the dominant contribution to \eqref{fthpl}. Using the integral representation of the Bessel function we obtain 

\begin{align}
    P_l(\cos(\theta))\to& \frac{1}{2\pi}\int_0^{2\pi} d\phi e^{i(2l+1)\sin(\theta/2) \cos\phi} \\
    =& \frac{1}{2\pi}\int_0^{2\pi} d\phi e^{i\vec{q} \cdot \vec{b}} 
\end{align}
where we have introduced the 2-vectors $\vec{q},\vec{b}$ defined by
\begin{equation}
    |\vec{q}|= 2\omega \sin(\theta/2)\,, \,\,\,b=|\vec{b}|=\frac{1}{\omega}(l+1/2)
\end{equation}
with $\phi$ as a relative angle. Thus we will identify $\vec{q}$ as the momentum transfer vector and $l+1/2=\omega b$ as the classical angular momentum in the orbital plane \cite{Andersson:2000tf}. So far no assumption has been made regarding $\omega$. In order to approximate the infinite sum over $l=1,2,\ldots$ by integration over $b$ we will take $\Delta b/b_c = \frac{1}{b_c\omega}\to 0$ for some characteristic impact parameter for which we expect the integral to localize. This high-energy regime is characteristic of the eikonal: as the wavelength is negligible with respect to the classical value of the impact parameter it leads to a point-particle behaviour. Note that this still overlaps with the Born approximation as long as $GM\omega \to 0$.

Putting everything together, \eqref{fthpl} can be written as 

\begin{equation}\label{ec:2eik}
   i f(\vec{q})=\frac{\omega}{2\pi}\int bdbd\phi (e^{i\chi(b)}-1)e^{i\vec{q}\cdot\vec{b}}\,,
\end{equation}
where the eikonal phase is given by
\begin{equation}
    \chi(b):=2\delta_l= 2\delta_{\omega b-1/2}
\end{equation}
Equivalently, we can write the inverse relation as

\begin{align}
  (e^{i\chi(b)}-1)=&\frac{i}{\omega }\int \frac{d^2\vec{q}}{(2\pi)}  f(\vec{q})e^{- i\vec{q}\cdot\vec{b}}\, \\
  =& i 8\pi M \int\frac{d^4 q}{(2\pi)^{2}}e^{iq\cdot b}\delta(2p_1 \cdot q)\delta(2k_2\cdot q) f(q)
\end{align}
where we have covariantized the first line by making use of the kinematics given in \cref{sec:2}. We can now implement the relation between QFT amplitudes and the classical NP scalars, $8\pi Mf(q)\to \langle A_{4}(q)\rangle $. Taking the leading Born order $GM\omega \to 0$ on both sides gives:

\begin{equation}
\chi_{\textrm{LO}}(b)=  \int\frac{d^4 q}{(2\pi)^2}e^{iq\cdot b}\delta(2p_1 \cdot q)\delta(2k_2\cdot q) \langle A_{4}(q)\rangle ,\label{eq:eikonal phase}
\end{equation}
 Notice that we have written the amplitude in the classical limit, i.e. $(s-M^2)^2\sim t M^2 \ll M^4$. However, as contact terms in $t=q^2$ will lead to ultralocal functions of the impact parameter $b$, it becomes clear that we can drop powers of $\textcolor{black}{M}^2t/(s-M^2)^2=\sin^2\theta/2$. This regime, i.e. 
  
  \begin{equation}\label{eq:Eikonal}
\textcolor{black}{M}^2 t \ll (s-M^2)^2 \ll M^4,
\end{equation}
  is precisely what we assumed to derive equation \eqref{ec:2eik}. The regime has been recently considered in the eikonal approximation  for massless particles scattering off heavy ones \cite{Bjerrum-Bohr:2016hpa,AccettulliHuber:2020oou}. Remarkably, it is contained as a limiting case of the classical prescription of Section \ref{sec:2}, i.e. $\theta\to 0$.
  
In order to illustrate the comparison to partial waves let us evaluate the eikonal amplitude \eqref{ec:2eik} using the leading approximation $\chi_{\textrm{LO}}$. This is computed from the tree-level amplitude \eqref{eq:eikonal phase}. As we review below in equation \eqref{eq:eikonal-phase-result}, in the Schwarzschild case ($a=0$) for any helicity of the wave, it is given by 

\begin{equation}
    \chi_{\rm{LO}}(b)=-4GM\omega \log\omega b +\Phi_{\textrm{IR}},
\end{equation}
where $\Phi_{\textrm{IR}}$ contains a $b$-independent infrared divergent phase. In our computation using partial waves, eq. \eqref{eq:scalar_amplitude_BHPT-Schw}, such a divergence is regulated by the form of asymptotic wavefunctions \eqref{eq:asymtpwa}, so we will ignore it here. Plugging this into \eqref{ec:2eik}, and assuming $b\ne 0$, the  integral evaluates to

\begin{equation}\label{newt}
    if(q) = i G M \frac{\Gamma[1-i\epsilon]}{\Gamma[1+i\epsilon]}\left(\sin(\theta/2)\right)^{-2+2i\epsilon}
\end{equation}
where we have  used $q^2= -4\sin^2(\theta/2)\omega^2$. This  result is precisely the Newtonian approximation obtained in \eqref{eq:scalar_amplitude_BHPT-Schw}, which dominates in the forward limit $\theta\to 0$ \cite{Andersson:2000tf}.
The fact that the leading eikonal can be used to derive the Newtonian phase was noted in \cite{Kabat:1992tb}, where it was used to further study the spectrum of bound states associated to the Newtonian potential. Bound states are characterized by the singularities of \eqref{newt} which appear at imaginary frequencies $\epsilon=i,2i,3i,\ldots$ as expected.

\subsection{Relation to Null Geodesics}

In order to derive observables from the eikonal approximation it is important to understand its direct relation to null geodesics in Kerr. This does not only reflect the universality of the optical limit of wave scattering, but also shows explicitly that it admits a point-particle description. Here we present an argument for the leading eikonal (tree-level), whereas in \cref{sec:subleading} we will check it holds to subleading (loop) orders.

In the harmonic gauge \eqref{eq:harmgauge}, we consider the momentum \textit{covector} $k_\mu= g_{\mu\nu}k^\nu$, where $k^\mu$ is a null four-velocity. Nicely, in terms of the covector the (linearized) geodesic equation for an affine parameter $\lambda$ takes a very simple form:

\begin{equation}
    \frac{dk_\mu}{d\lambda}= \frac{\kappa}{2}\partial_{\mu}h_{\alpha \beta}(x) k^\alpha k^\beta \,.
\end{equation}
Interpreting $k_\mu$ as canonical momenta, we can already see from this form that the contraction $h_{\alpha \beta}(x) k^\alpha k^\beta$ corresponds to an effective potential, and in fact is nothing but the four-point amplitude. Let us be more explicit. Integrating this equation along the trajectory $x^\mu (\lambda)=k_2^\mu \lambda + b^\mu +\mathcal{O}(G)$ (at the linearized level we ignore the Coulomb drag) we obtain

\begin{equation}\label{eq:geod}
  k_2^\mu(+\infty) -k_2^\mu(-\infty)= \Delta k^\mu= \frac{\kappa}{2}\int d\lambda \partial^{\mu}h_{\alpha \beta}(k_2 \lambda + b) k^\alpha_2 k^\beta_2 \,.
\end{equation}
Note that at $\lambda\to \pm \infty$ we have raised the indices of $k_\mu$ using the flat metric $\eta^{\mu\nu}$. This is because geodesics reach null infinity where the metric is (asymptotically) flat, hence we can identify $k_\mu$ with the momentum vector in the scattering amplitude.

We now plug in the Fourier representation of the linearized gravitational field generated by a source with momentum $p_1$  (cf. equations \eqref{eq:justin position }-\eqref{eq:justin vertex classical}) 

\begin{equation}
    h_{\mu \nu}(x)= \int \frac{d^{4}q}{(2\pi)^3} \frac{\delta(2q\cdot p_1)}{(q^0+i\epsilon)^2-\vec{q}^2}e^{iq\cdot x}\hat{h}_{\mu \nu}(q)\,.
\end{equation}
Notice we have introduced explicitly the $i\epsilon$ prescription for the retarded propagator, as naturally arising from classical computation. 
Recall that the field component $\hat{h}_{\mu \nu}$ (which can be read off from \eqref{eq:justin vertex classical} in the case of the Kerr metric) has classical scaling and can be interpreted as an off-shell three-point vertex including the harmonic projector. From \eqref{eq:geod} we obtain the momentum deflection
\begin{eqnarray}\label{eq:deltapgeo}
   \Delta k^\mu&=&\textcolor{black}{ \frac{\kappa}{2}}\int\frac{ d^{4}q}{(2\pi)^3} \frac{\delta(2q\cdot p_1)}{(q^0+i\epsilon)^2-\vec{q}^2}q^{\mu }\hat{h}_{\alpha\beta}(q)k^\alpha_2 k^\beta_2 \int d\lambda e^{iq\cdot (k_2 \lambda + b)}  \nonumber\,, \\
   &=&\kappa \int \frac{d^{4}q}{(2\pi)^2} e^{iq\cdot b} \delta(2q\cdot p_1)\delta(2q\cdot k_2)q^\mu \frac{\hat{h}_{\alpha\beta}(q)k^\alpha_2 k^\beta_2 }{q^2+i\epsilon}   \,,
\end{eqnarray}
where in the second line we switched from the retarded to the Feynman $i\epsilon$-prescription for the massless propagator using the support of the on-shell delta functions (see \cref{sec:eikonal-scalars}). 
As we explained in section \cref{sec:relsch}, the tensor $k_2^{\alpha}k_2^{\beta}$ corresponds to the massless three-point vertex for a scalar emitting a graviton. Moreover, it turns out that in the on-shell limit $q^2\to 0$ this is not only the scalar-graviton amplitude but also the 3-pt amplitude for any massless particle emitting a graviton, up to a phase. More precisely, note that from gauge invariance the amplitude for a massless scalar emitting a graviton is simply 

\begin{equation}
    \langle A_3^{0,\mu \nu}\rangle \propto  k_2^\mu k_2^\nu\,.
\end{equation}
Furthermore, as in general the three-point amplitude is completely fixed by helicity weights, it is easy to see that the case $h\neq 0$ simply yields an helicity factor

\begin{equation}
   \langle A_3^{h,-h,\mu \nu}\rangle \propto  k_2^\mu k_2^\nu (-i\mathcal{N}^X)^{2h}\,.
\end{equation}
where $(-i\mathcal{N}^X)^{2h}$ turns into a phase when evaluated in the real momenta of the 4-pt amplitude. It can be absorbed by a little-group transformation of the massless states, setting $-i\mathcal{N}^X=1$. We show this explicitly in Appendix \cref{genhel}, with three-point amplitudes given in \eqref{eq:3ptmassls} in terms of spinor-helicity variables (parametrized such that $-i\mathcal{N}^X=1$). Then, as $q^2\to0$ we can replace the 4-pt amplitude for any helicity scattering by its 3-pt factorization:

\begin{equation}\label{repa4h}
    \langle A_4^h(q) \rangle \to \kappa \frac{\hat{h}_{\alpha\beta}(q)k^\alpha_2 k^\beta_2 }{q^2+i\epsilon} 
\end{equation}
where the fact that the RHS is independent of the helicity confirms that the $q^2\to 0$ limit of $\langle A_4^h(q) \rangle \,$ is indeed universal, see also \cref{genhel}.

Moreover, using the fact that we only require the leading term in $q^2$ in the integrand of \eqref{eq:deltapgeo}, we use the replacement \eqref{repa4h} and obtain
\begin{eqnarray}\label{eq:deltapgeo-final}
   \Delta k^\mu 
   &=& \int \frac{d^{4}q}{(2\pi)^2} e^{iq\cdot b} \delta(2q\cdot p_1)\delta(2q\cdot k_2)q^\mu \langle A_4^h(q) \rangle \,.
\end{eqnarray}
which is nothing but the massless analog of the KMO formula derived from QFT in \cite{Kosower:2018adc}. Here we have derived it from the geodesic equation, showing that the eikonal limit of wave scattering is directly related to particle motion. More precisely, the latter the equation can be written as

\begin{equation}\label{thetatochi}
    \Delta k^\mu = \frac{\partial}{\partial b_\mu} \chi(b),
\end{equation}
where we have now used the eikonal phase \eqref{eq:eikonal phase}. This means that the eikonal phase can be used to derive classical observables associated to a null geodesic. In particular, as the momentum deflects in the direction of the impact parameter (at leading order) the previous formula is equivalent to

\begin{equation}
    -\theta = \frac{ \hat{b}^\mu \Delta k_{\mu}}{|\vec{k}|} =\frac{1}{\mathcal{\omega}}\frac{\partial}{\partial b}\chi =\frac{\partial}{\partial J}\chi,\label{eq:scattering-angle-eikonal}
\end{equation}
where we used $\hat{b}^\mu = \frac{\partial b^\mu}{\partial b}$. Here $\theta$ is the classical (eikonal) scattering angle characterizing the scattering of the particle off the Kerr background.  In the last equation, $J=\omega b$ is the orbital angular momentum of the particle.

Equation \eqref{thetatochi} reveals that $\chi(b)$ is naturally interpreted as the radial action part of the action of a \textit{null geodesic}. It corresponds to the Hamilton-Jacobi equation for endpoint observables (or in QFT terms to classical the saddle point of \eqref{fthpl}). As such, it holds up to subleading eikonal orders as we will check explicitly by comparing to both geodesics in Kerr and QFT results. This contrasts with the massive eikonal case as studied recently \cite{Kalin:2020fhe} where the radial action is not a geodesic action but instead describes the effective relative motion of the two bodies.\footnote{In the massless limit of the background, i.e. boosted Schwarzschild or Kerr metric, the correspondence to null geodesics is well known at least at leading eikonal order, see e.g. \cite{tHooft:1987vrq,Amati:1987uf,Camanho:2014apa,Cristofoli:2020hnk}}.
  
Following Hamilton-Jacobi theory, the conjugate equation to \eqref{eq:scattering-angle-eikonal} gives the Shapiro time delay associated to the null trajectory:
\begin{align}
t_S & =\frac{\partial}{\partial\mathcal{\omega}}\chi\label{eq:time-delay-eikonal}.
\end{align}

Next we carry out explicitly the leading eikonal approximation from our previous QFT amplitudes with spin, and obtain the corresponding observables in the polar and equatorial configurations.

\subsection{Leading Eikonal in Kerr}\label{sec:eikonal-scalars}

We will illustrate the computation with a scalar wave. That this leads to the same result independently of helicities is concluded in \cref{genhel}. Consider then the scalar amplitude evaluated in the kinematics \eqref{eq:kinematics} 

\begin{equation}\label{scalar-kerr2}
    \langle A_{4}^{S\to \infty ,h=0} \rangle=\frac{\kappa^2 M^2\omega}{ q^2}\left[\omega \cosh{a{\cdot}q}+ i  \vec{a}{\cdot}\vec{q}{\times}\vec{k}_2 \frac{\sinh{a{\cdot}q}}{a{\cdot}q} \right], 
\end{equation}
in the $t\to 0$ regime. A way to treat the above limit is precisely to take $\theta\to 0$ in the parametrization \eqref{eq:kinematics}-\eqref{eq:polar spin}. This leads to

\begin{equation}
    \vec{a}{\cdot}\vec{q}{\times}\vec{k}_2 \to -\omega \theta a_y \,,\quad a\cdot q\to  \omega \theta a_x\,.
\end{equation}
and thus matches the universal form \eqref{eq:eikoApp} as promised. However, in order to evaluate the eikonal integral \eqref{eq:eikonal phase} we should first massage the expression using the support of $t\to 0$. To do this, note
that the four-point amplitude behaves as $1/t$ times the residue, which we can evaluate at the strict $t=0$ limit. We thus need to pick a branch since $t=  \braket{23}[32]$.  Each choice of vanishing brackets
represents the exchange of a graviton of opposite  helicity and as such are related by parity conjugation.
Moreover, on the support of the $t-$channel residues, the argument of the trigonometric functions in \eqref{scalar-kerr2} becomes
\begin{align}
 q{\cdot}a\big|_{[23]=0}& = i \frac{\epsilon(q,a,p_1,k_2)}{p_1\cdot k_2}= i\vec{q}{\cdot}\hat{k}_2\times\vec{a}\\
q{\cdot}a\big|_{\braket{23}=0}&=-i \frac{\epsilon(q,a,p_1,k_2)}{p_1\cdot k_2}=-i\vec{q}{\cdot}\hat{k}_2\times\vec{a},
\end{align}
After inserting this into \eqref{scalar-kerr2} we observe that both residues give the same contribution, and thus we can choose either branch. With this in mind, and  restoring the overall factor of $i$ in the amplitude we can write
\begin{equation}\label{eq:eikonal-scalar}
\langle A_{4}^{K,h=0}\rangle=\frac{i\kappa^{2}\left(M\mathcal{\omega}\right)^{2}}{q^{2}+i\epsilon} e^{i\vec{q}{\cdot}\hat{k}_2\times\vec{a}}+\mathcal{O}(t^0)\,.
\end{equation}
As expected, we note that the $h=0$ amplitude is invariant under the helicity flip (i.e. time reversal) operation introduced in  \cref{sec:polapp}, namely by taking $\langle A_{4}^{K,h=0}\rangle \to \langle A_{4}^{K,h=0}\rangle^*$ and $a\to -a$. However, such invariance is not preserved once we compute the eikonal phase which selects a time direction, i.e. we define our incoming momenta.\\

The integral $(\ref{eq:eikonal phase})$ has been done explicitly in \cite{Guevara:2019fsj}, and we recap its derivation here. First we split the 4-dimensional momentum transfer into the longitudinal and transverse directions as 
\begin{equation}
    q=\alpha_1 \, p_1 +\alpha_2\, k_2 +q_\perp, 
\end{equation}
Next we can use the delta functions to do the integrals in the time and  longitudinal direction, which translate into taking $\alpha_1=\alpha_2=0$, independent of the $i\epsilon$-prescription taken. In addition, using $p_1{\cdot}k_2=M\omega$, and replacing \eqref{eq:eikonal-scalar} into \eqref{eq:eikonal phase}, we arrive at the integral
\begin{equation}
\chi(b)=8\pi G M\omega \int\frac{d^{2}q_\perp}{\left(2\pi\right)^{2}}\frac{e^{i\vec{q}_\perp{\cdot}(\vec{b}+\hat{k}_2\times \vec{a})}}{\vec{q}_\perp^2} ,
\end{equation}
where we have removed the $i\epsilon$ factor since we are left with a well behaved integral. We can explicitly  evaluate it to get
\begin{equation}\label{eq:eikonal-phase-result}
\chi(b)=-4GM\omega
\log|\omega \vec{b}+ \vec{k}_2\times\vec{a}|+\Phi_{\textrm{IR}}\,.
\end{equation}
Here  we have performed the 2d integration noting that $\vec{b}$ and $\vec{k}_2\times \vec{a}$ are both orthogonal to $p_1$ and $k_2$. The IR piece is independent of both the impact parameter and the spin, and can be given as
\begin{equation}
  \Phi_{\textrm{IR}}= 4GM\omega\log\omega/\mu= -4GM\omega\left[\frac{1}{2\varepsilon}+\frac{\gamma_E}{2} -\log\omega \right]
\end{equation}
where in the second form we used dimensional regularization with the dim-reg parameter $\varepsilon$. This was expected since the massless amplitude is known to contain an IR divergent phase.\footnote{Indeed, eikonal methods have proven powerful in evaluating the IR-divergent phase for massless amplitudes, see e.g. \cite{Naculich:2011ry}.}

We can make contact with the results of the previous section, based on BHPT, as follows. Using the coordinate system described around eq. \eqref{Eq:fSpher} we can translate the eikonal phase into the form
\begin{equation}\label{eq:eikonal-phase-coord}
\chi(b,m)-\Phi_{\textrm{IR}}=-2GM\omega
\log\left[ b^2\omega^2+a^2\omega^2\sin^2\gamma + 2 a\omega m \right]= 2\,\delta_{l m}^{\textrm{eik}}(\gamma)
\end{equation}
where we recall the orbital angular momentum is $(b \omega)^2=l(l+1)\approx(l+1/2)^2$ and the azimuthal component (in the direction of $\vec{a}$) is given by $m$. In deriving this result we have used that the orbital angular momentum vector reads $\vec{J}=\vec{k}_2 \times \vec{b}$. The result can then be compared to the results in \eqref{Eq:fSpher} by writing
\begin{equation}
    f_{lm}(\gamma)=\frac{2\pi}{i\omega}Y_{lm}(0,0)(e^{2i\delta_{lm}(\gamma)}-1)\,,
\end{equation}
and then taking $l\to\infty$ in $\delta_{lm}$, with $m/l$ fixed. Performing the $l\to \infty, \omega\to \infty$ (note that the anomalous terms do not contribute in this regime) expansion of the results in section \cref{sec:Hos} we have managed to check the agreement directly for the cases of equatorial ($m=l,\gamma=\pi/2$) and polar ($m=0,\gamma=0$) configurations. However, we will provide an indirect check of the phases by comparing observables in the next section.

\subsection{Towards observables from the spinning eikonal phase}\label{sec:subleading}

In the spinless (Schwarzschild) case we have seen that the QFT eikonal phase leads to observables such as the scattering angle $\theta$ and the time delay $\Delta t$. We now extend this correspondence to the spinning case, as well as to higher loop cases for the case of scalar black holes.

In the spinning case, the definition of the scattering angle must be supplemented since classical dynamics do not take place on a plane. Instead, we have two different angular deviations corresponding to polar $\Delta \theta$ and azimuthal $\Delta \phi$ coordinates. Interestingly, both deviations can be encoded in the phases and can be obtained from the Teukolsky equation or from the QFT eikonal.

To illustrate this we will analyze two configurations: polar and equatorial scattering. Let us start with \textit{equatorial scattering}, for which the incoming wave impinges on the equatorial plane of the Kerr BH, this implies $\vec{k_2}\perp \vec{a}$ and also $\vec{b}\perp \vec{a}$. Thus, this is nothing but the `aligned-spin' setup $\vec{J}//\vec{a}$ considered in \cite{Vines:2017hyw,Guevara:2018wpp}. Now, in the language of the BHPT phases \eqref{Eq:fSpher}, we see that $\vec{k_2}\perp \vec{a}$ implies $\gamma=\pi/2$ and $\vec{J}//\vec{a}$ implies $m=l$ (recall $m$ is the azimuthal component of $\vec{J}$). The phase \eqref{eq:eikonal-phase-coord} (without the IR piece) becomes simply

\begin{equation}
    \chi(b,m=b\omega)=-4GM\omega\log|b+a|
\end{equation}

As anticipated, it can be shown that this is nothing but the radial piece of the on-shell action for an equatorial null geodesic in the Kerr metric, 
\begin{equation}
    \oint p_{r}dr=2\int_{r_{\min}}^{\infty}\frac{\omega rdr}{r^{2}+a^{2}-2GMr}\sqrt{r^{2}+a^{2}-b^{2}+\frac{2GM}{r}(b-a)^{2}}\label{null geode}
\end{equation}
expanded at leading order in $GM\omega$. The radial action is computed by identifying two conserved charges of geodesic motion; the energy $\omega$ and the azimuthal angular momentum $\omega b$. In fact, due to the spin-alignment the scattering does take place on a plane, as in the spinless case. Moreover, this observation was used in \cite{Guevara:2018wpp} as a motivation to compute the first correction of the eikonal phase from 1-loop amplitudes in the massive case. Its massless limit reads:
\begin{equation}
     \chi_{\textrm{1-loop}}(b,m=b\omega)=-\frac{\pi(GM)^{2}\omega}{2a^{2}}\left(b-4a-\frac{(b-a)^{4}}{(b^{2}-a^{2})^{3/2}}\right)\,,
\end{equation}
which can be seen to agree with the subleading order of the null geodesic action \eqref{null geode}. In fact, we have checked explicitly that it agrees with the     BHPT phases $\delta_{l,m=l}(\pi/2)$ computed in \cref{sec:Hos} up to subleading eikonal order and up to order $a^5$. Thus, this provides a crosscheck of both the BHPT phases and the QFT amplitudes up to 1-loop, as compared to classical null geodesics. One can proceed in the same fashion in order to compare higher-loop amplitudes with subleading phases and with the expanded radial action. For the spinless case, we have checked such agreement up to subsubleading order $\sim GM\omega(GM/b)^2$: From the results of \cref{sec:Hos}, using $\omega\to \infty, b\to \infty$ one finds 
\begin{equation}
     \chi(b,m=b\omega)=-4   GM\omega\left( \log b-\frac{15\pi GM}{16 b} -\frac{16G^2M^2}{3b^2}+\ldots\right) +\mathcal{O}(a)\,,
\end{equation}
which agrees with the probe limit result (6.29) of  \cite{DiVecchia:2021bdo}, in the massless limit. 

To compute the deflection angle $\Delta \theta$ we simply note that it is conjugate to the azimuthal angular momentum $m$, thus they are linked via the action \eqref{null geode} as
\begin{equation}
    \Delta \theta =\frac{1}{\omega} \frac{\partial \chi(b,m=b\omega)}{\partial b}= -\frac{4GM}{b+ a}\,,
\end{equation}
which agrees with the ultrarelativistic ($v\to 1$) limit of the massive scalar result \cite{Guevara:2018wpp} as expected.\footnote{It differs, however, from the boosted Kerr result derived in \cite{Cristofoli:2020hnk} via similar methods.} A similar matching holds up to 1-loop.

Let us now focus on the more interesting case of \textit{polar} configuration, which we have touched upon in the previous sections. In this case the wave impinges on the spin axis, i.e. $\gamma=0$. Because of this $\vec{J}$ is also perpendicular to $\vec{a}$, thus $m=0$. This automatically implies that the phases \eqref{eq:eikonal-phase-coord} do not depend on the spin and reduce to the scalar case. We have checked this agreement explicitly between the QFT eikonal phase and the BHPT phases. Indeed, we anticipated this from the fact that the spin contributions to the series \eqref{Eq:fSpher} truncate at a finite value $l_{\textrm{max}}$ for the polar configuration, as mentioned. On the other hand, computing an effective classical action for polar null geodesics is a more difficult task, partly due to the fact that scattering is not planar and there is  a dynamical angular component $p_\theta$ in addition to the radial piece $p_r$.\footnote{Using the WKB approximation a closed form has been proposed, however, in \cite{Glampedakis:2001cx}.}

Instead, let us introduce the connection between the phase and the scattering angles, both of which are now non-trivial:
\begin{align}
    \Delta\theta=2\frac{\partial \delta^{\textrm{eik}}_{lm}(\gamma=0)}{\partial l}\,\,\,,\qquad  \Delta\phi=2\frac{\partial \delta^{\textrm{eik}}_{lm}(\gamma=0)}{\partial m}  \label{eq:obspolar}
\end{align}
which are then evaluated at $m=0$. Thus, even though the effective action $\chi(b,0)$ does not depend on spin, its derivatives at $m=0$, corresponding to the associated observables, do introduce spin corrections. The relation for $\Delta\phi$ is expected from Hamilton-Jacobi theory since the azimuthal component $m$ is conserved (due to axial symmetry) and conjugate to $\phi$. The relation for $\Delta \theta$ is more mysterious since the orbital angular momentum is not conserved. However, it can be seen that in this case, the \textit{initial} orbital angular momentum $\vec{J}=\vec{k}_2\times \vec{b}$ is indeed conjugate to $\theta$.

We have checked the agreement in \eqref{eq:obspolar} between the BHPT phases and the leading eikonal phases \eqref{eq:eikonal-phase-coord}. Moreover, using the results of \cite{Kosmopoulos:2021zoq,Liu:2021zxr} for the subleading eikonal, or 1-loop amplitudes up to order $\mathcal{O}(a^2)$, we have computed both $\Delta\theta$ and $\Delta\phi$ and find perfect agreement with the BHPT phases of \cref{sec:Hos}. Furthermore, up to this same 1-loop order both angles are in agreement with an independent derivation based on polar null geodesics in a Kerr background. This provides an important verification of the validity of formulae \eqref{eq:obspolar} for polar scattering. We leave the extension thereof to generic configurations of $\gamma$ and $m$ for future investigation.\footnote{Both equatorial and polar configurations can be understood as particular cases of the singular configuration $b\cdot a=0$, namely $\sin\gamma\approx m/l$. It would be interesting to explore this configuration in more detail.}

To close this section, we also evaluate the time delay \eqref{eq:time-delay-eikonal}. Let us focus on the aligned-spin case for simplicity. Physically, only time differences can be measured. Following e.g. \cite{AccettulliHuber:2020oou,Afkhami-Jeddi:2018apj} we choose the difference between the time measured by an observer at $b_0$ and that measured by another at a much larger position $b$. With this in mind we get
\begin{equation}\label{eq:deltat}
    \Delta t_S = 4GM\log\left|\frac{b+ a}{b_0+ a} \right|.
\end{equation}
This is also an universal result, and setting $a=0$  recovers the time delay for the scattering of a gravitational wave off a Schwarzschild Black Hole at leading order in $G$ \cite{AccettulliHuber:2020oou}.
 Importantly we notice that $\Delta t_S$ in \eqref{eq:deltat} is positive at large $b>b_0$, reflecting no causality violation to this order of perturbation theory \footnote{The opposite, small $b$, limit can lead to causality violation as shown in \cite{Afkhami-Jeddi:2018apj} for massive spinning particles, which however lies outside the domain of validity of our approximation \eqref{eq:eikonal phase}. We thank R. Roiban for this observation. 
 } A positive time delay is in fact a strong signature of unitarity \cite{Adams:2006sv}.

\section{Discussion}\label{sec:discussion}
We have proposed a novel QFT picture of low energy perturbations of black hole geometries, realized as the scattering of waves, together with the corresponding observables.

We have worked in a point-particle limit. In the Schwarzschild case it should be noted that as $\lambda$ is much bigger than the size of the black hole its internal structure may not be probed. Indeed, at leading order in $GM\omega$, the classical (tree-level) scattering amplitude is independent of the nature of the compact object, whether it corresponds to a black hole or other spherical body
in \cite{PhysRevD.13.775}. This is a reflection of an underlying equivalence principle: In the QFT setup this reflects the universality of Weinberg's soft theorem \cite{Weinberg:1965nx}. In fact, it continues to be true for spinning objects at \textit{linear} order in their spins when scattering with gravitons, as dictated by the universal subleading soft-theorem \cite{Cachazo:2014fwa}. However, such behaviour breaks down at higher orders in spin, or higher multipoles, \cite{Guevara:2018wpp,Chung:2018kqs,Bautista:2019evw}. This is crucial: For higher orders in the spin, entering through long-wavelength corrections $\sim a\omega$ the Kerr BH should be probed and distinguished from other compact objects by means of its unique spin-multipole structure.  

In this work we have provided a precise prescription for approaching the point-particle limit of solutions to the Teukolsky equation. For $h=0$, it is motivated by the fact that the Klein-Gordon equation can be solved to all orders in spin for the point-particle (Born) approximation $GM\omega\to 0$, and that we can recover such result from BHPT if we analytically extend its solutions to $a/GM\to\infty$. Additionally, extra contributions containing polygamma functions of $a$ emerge at low values of the angular momentum $l$. We have identified such contributions as contact terms which were absent in the Born amplitude and hence it is natural to discard them, but further investigation is required to understand their role.

We have encountered dissipative effects associated, through unitarity, to the imaginary part of the scattering phases. Even though we have discarded them, as they are not expected for the tree-level S-Matrix, it would be interesting to match the imaginary piece with a higher-loop computation, as has been done recently in the context of the 2-body problem \cite{Herrmann:2021tct,DiVecchia:2021bdo}, or with the EFT approach to black-hole absorption exemplified by \cite{Goldberger:2020fot}.

There are many intriguing future extensions of this construction. We expect the equivalence presented here to have wide applications in classical black-hole physics and not only for the case of wave scattering. For instance, at higher perturbative orders it may help to elucidate diverse results that seem miraculous from a classical perspective, such as the Kerr/CFT duality, integrability, and the associated hidden symmetries, see e.g. \cite{Castro:2010fd}.

\appendix

\acknowledgments
We thank Zvi Bern, Shahar Hadar, Alok Laddha, Alex Lupsasca,  Donal O' Connell, Julio Parra-Martinez, Andrew Strominger, Shan-ming Ruan and Maarten van de Meent for useful discussions. We are grateful of Donal O'Connell for sharing a preliminary draft of \cite{CGKO}. A.G. is supported by a Junior Fellowship at the Harvard Society of Fellows, as well as by the DOE grant de-sc/0007870. Y.F.B.  acknowledges the Natural Sciences and Engineering Research Council of Canada (NSERC) the financial support.  Research at Perimeter Institute is supported by the Government of Canada through the Department of Innovation, Science and Economic Development Canada and by the Province of Ontario through the
Ministry of Research, Innovation and Science.

\section{Teukolsky equation}
\label{Teuk-App}
%. 
In this appendix we will overview relevant results in the analytic approach to solving the Teukolsky equation. For more details we direct the reader to the review \cite{Sasaki:2003xr}.
\begin{center}
\begin{table}
\begin{centering}
\begin{tabular}{|c|c|c|}
\hline 
Teukolsky scalar $\psi$ & Tetrad Representation & Helicity $h$\tabularnewline
\hline 
\hline 
$\phi$ &  & $0$\tabularnewline
\hline 
$\Phi_{0}$ & $F_{\mu\nu}l^{\mu}m^{\nu}$ & $+1$\tabularnewline
\hline 
$\rho^{-1}\Phi_{2}$ & $\frac{\rho^{-1}}{2}F_{\mu\nu}(l^{\mu}n^{\nu}+\bar{m}^{\mu}m^{\nu})$ & $-1$\tabularnewline
\hline 
$\Psi_{0}$ & $-C_{\mu\nu\rho\sigma}l^{\mu}m^{\nu}l^{\rho}m^{\sigma}$ & $+2$\tabularnewline
\hline 
$\rho^{-4}\Psi_{4}$ & $-\rho^{-4}C_{\mu\nu\rho\sigma}n^{\mu}\bar{m}^{\nu}n^{\rho}\bar{m}^{\sigma}$ & $-2$\tabularnewline
\hline 
\end{tabular}
\par\end{centering}
\caption{Teukolsky scalar for perturbations of helicity $h$. Here $F_{\mu\nu}$ and $C_{\mu\nu\rho\sigma}$ correspond to the Faraday and Weyl tensors respectively, whereas the  spin coefficient $\rho^{-1}=(r-ia \cos\theta)$. }
\label{tab:scalar}
\end{table}
\par\end{center}
For vacuum perturbations of the Kerr BH, and in Boyer-Linquist coordinates, the Teukolsky scalar $\psi$ (See \cref{tab:scalar} ), 
satisfies the homogeneous Teukolsky equation \cite{Teukolsky:1972my}
\begin{align}
&\left[\frac{(r^2+a^2)^2}{\Delta}-a^2\sin^2\theta\right]\spdiff{\psi}{t}+\frac{4 M a r}{\Delta}\frac{\partial^2 \psi}{\partial t \partial \varphi}+
\left[\frac{a^2}{\Delta}-\frac{1}{\sin^2\theta}\right]\spdiff{\psi}{\varphi} 
-\Delta^{-h}\pdiff{}{r}\left(\Delta^{h+1}\pdiff{\psi}{r}\right) \nonumber\\
&-\frac{1}{\sin \theta}\pdiff{}{\theta}\left(\sin\theta \pdiff{\psi}{\theta}\right)  
-2h\left[\frac{a(r-M)}{\Delta}+\frac{i \cos\theta}{\sin^2\theta}\right]\pdiff{\psi}{\varphi}
-2 h \left[\frac{M(r^2-a^2)}{\Delta}-r -i a \cos\theta\right]\pdiff{\psi}{t}\nonumber \\
&\qquad\qquad+h(h\cot^2\theta-1)\psi = 0,
\label{Eq:TeukMaster}
\end{align} 
with   $h$ the helicity of the perturbation\footnote{Notice the more standard BHPT notation for labeling the spin/helicicity of the wave perturbation is $s$. Here we have used $h$ in order to keep the same notation used in the main body of the paper. }.
This is separable in the frequency domain as 
\begin{align}
	\psi(t,r,\theta,\phi)=\sum_{\ell m}\int d\omega e^{-i \omega t}{}_{h}Z_{l m \omega}\Rlm{h}{\ell}{m}{\omega}(r)\sSlm{h}{\ell}{m}(\theta,\phi,a \omega).
\end{align}
Here, ${}_{h}Z_{l m \omega}$ are normalization coefficients.  $\Rlm{h}{\ell}{m}{\omega}(r)$ are solutions to the homogeneous radial Teukolsky equation and $\sSlm{h}{\ell}{m}(\theta,\phi,a \omega)$ are the spin-weighted spheroidal harmonics with respective defining equations
\begin{align}
\bigg[\Delta^{-h}\diff{}{r}\left(\Delta^{h+1}\diff{}{r}\right)+&\frac{K^2-2 i h (r-M)K}{\Delta}
+4 i h \omega r -{}_h \lambda_{lm}\bigg]{}_h R_{\ell m \omega}(r)
=0, \label{Eq:RadTeuk} 
\end{align}
and
\begin{align}
\bigg[\frac{1}{\sin \theta}\diff{}{\theta}\left(\sin\theta \diff{  }{\theta}\right)-&a^2 \omega^2 \sin^2\theta
-\frac{(m+h\cos\theta)^2}{\sin^2\theta} \nonumber \\
&-2 a \omega h \cos\theta+h+2 m a \omega + {}_h \lambda_{lm}\bigg]\sSlm h l m (\theta, \varphi; a \omega)=0. \label{Eq:Slmeq} 
\end{align}
where $K = (r^2+a^2)\omega - a m$, and 
${}_h \lambda_{lm}$ is the spheroidal eigenvalue.  For the Schwarzschild limit they reduce to the eigenvalues of the spherical harmonics $\lambda\to \ell(\ell+1)$.

For our needs we will require a vacuum Teukolsky solution, typically labelled $\Rin{h}{\ell}{m}{\omega}$, which satisfies the physical boundary condition of purely ingoing waves at the horizon, namely \cite{futterman88}
\begin{align}
	\Rin{h}{\ell}{m}{\omega}(r)=\Btrans{\ell}{m}{\omega} \Delta^{-h}e^{-i\tilde{\omega} \rs}, \qquad r\rightarrow r_+,
\end{align}
where $r_+=M+\sqrt{M^2-a^2}$ is the location of the outer horizon, $\tilde{\omega}=\omega-\frac{m a}{2 M r_+}$, and $\Btrans{\ell}{m}{\omega}$ is the so called transmission coefficient, which is understood to be a function of $h$.
Imposing this boundary condition fixes the asymptotic form at radial infinity for each $\ell,m$ mode to be  \cite{futterman88}
\begin{align}
	\Rin{h}{\ell}{m}{\omega}(r)=\Binc{\ell}{m}{\omega} r^{-1}e^{-i\omega \rs}+\Bref{\ell}{m}{\omega} r^{-(2h+1)} e^{i\omega \rs} , \qquad r\rightarrow \infty, \label{Eq:InfFormTeuk}
\end{align}
where $\Binc{\ell}{m}{\omega}$ and $\Bref{\ell}{m}{\omega}$ are the incident and reflection coefficients also functions of $h$.
Solutions to the radial Teukosky equation can  be written as infinite series of hypergeometric functions or confluent hypergeometric functions, depending on the required asymptotic boundary conditions \cite{Leaver86_1,Leaver86_2,Mino:1996nk,Sasaki:2003xr}. Investigation of the asymptotic behaviour of these infinite series yields expressions for the incident and reflection coefficients:
\begin{align}
\Binc{\ell}{m}{\omega}
=&\omega^{-1}\left[{K}_{\nu}-
ie^{-i\pi\nu} \frac{\sin \pi(\nu-h+i\epsilon)}
{\sin \pi(\nu+h-i\epsilon)}
{K}_{-\nu-1}\right]A_{+}^{\nu} e^{-i(\epsilon\ln\epsilon -\frac{1-\kappa}{2}\epsilon)},\label{Eq:Binc}
\\
\Bref{\ell}{m}{\omega}
=&\omega^{-1-2h}\left[{K}_{\nu}
+ie^{i\pi\nu} {K}_{-\nu-1}\right]A_{-}^{\nu}
e^{i(\epsilon\ln\epsilon -\frac{1-\kappa}{2}\epsilon)},\label{Eq:Bref}
\end{align}
with
\begin{align}
&A_{+}^\nu=e^{-{\pi\over 2}\epsilon}e^{{\pi\over 2}i(\nu+1-h)}
2^{-1+h-i\epsilon}{\Gamma(\nu+1-h+i\epsilon)\over 
\Gamma(\nu+1+h-i\epsilon)}\sum_{n=-\infty}^{+\infty}a_n^\nu,\\
&A_{-}^\nu=2^{-1-h+i\epsilon}e^{-{\pi\over 2}i(\nu+1+h)}e^{-{\pi\over 2}\epsilon}
\sum_{n=-\infty}^{+\infty}(-1)^n{(\nu+1+h-i\epsilon)_n\over 
(\nu+1-h+i\epsilon)_n}a_n^\nu, 
\end{align}
and 
\begin{align}
K_{\nu}=&	\frac{e^{i\epsilon\kappa}(2\epsilon \kappa )^{h-\nu-r}2^{-h}i^{r}
	\Gamma(1-h-2i\epsilon_+)\Gamma(r+2\nu+2)}
	{\Gamma(r+\nu+1-h+i\epsilon)
	\Gamma(r+\nu+1+i\tau)\Gamma(r+\nu+1+h+i\epsilon)}
	\nonumber\\
	&\times \left ( \sum_{n=r}^{\infty}
	(-1)^n\, \frac{\Gamma(n+r+2\nu+1)}{(n-r)!}
	\frac{\Gamma(n+\nu+1+h+i\epsilon)}{\Gamma(n+\nu+1-h-i\epsilon)}
	\frac{\Gamma(n+\nu+1+i\tau)}{\Gamma(n+\nu+1-i\tau)}
	\,a_n^{\nu}\right)
	\nonumber\\
	&\times \left(\sum_{n=-\infty}^{r}
	\frac{(-1)^n}{(r-n)!
	(r+2\nu+2)_n}\frac{(\nu+1+h-i\epsilon)_n}{(\nu+1-h+i\epsilon)_n}
	a_n^{\nu}\right)^{-1}.
	\label{eq:Knu}
\end{align}
Here $r$ is a free parameter (not to be confused with the radial co-ordinate) we set to be $0$, $\epsilon=2 G M \omega$, $\kappa=\sqrt{1-a^{\star 2}}$, $a^\star=\frac{a}{GM}$, $\tau=\frac{\epsilon-m a^{\star}}{\kappa}$ and $ \epsilon_{\pm}=\frac{\epsilon\pm\tau}{2}$. In these expressions the series coefficients $a_n^{\nu}$ satisfy 3 term recurrence relations and the `renormalised angular momentum' $\nu$ is determined by insisting the series all converge. 

While complicated, calculating the low frequency expansions of $\Binc{\ell}{m}{\omega}$ and $\Binc{\ell}{m}{\omega}$ ultimately come down to determining low frequency expansions of $a_n^{\nu}$ and $\nu$. These have been extensively studied (see e.g. \cite{Sasaki:2003xr,Kavanagh:2016idg}), and so we will not discuss this problem here. The relevant results will soon be available in the Black Hole Perturbation Toolkit \cite{BHPToolkit}.

\section{Eikonal Amplitudes: Universality for  general helicity}\label{genhel}

In this appendix we show the universality of the eikonal phase  mentioned in \cref{sec:eikonal-scalars}.  
Since for the eikonal limit only the $t$-channel residues are important, we evaluate the 4-pt amplitude for waves of general helicity $h$, using t-channel gluing. This has been done for the case of Schwarzschild in \cite{Bjerrum-Bohr:2016hpa} and it is interesting how such universality extends non-trivially for Kerr, of course reflecting the underlying equivalence principle.

We have two kinds of scattering processes, in the first one the helicity of the wave is conserved, whereas in the second one the helicity is flipped. As we will see shortly, the  latter is subleading in the forward/eikonal limit and therefore only the former will be relevant for the computation of physical observables from the eikonal phase.

The four-point amplitude can be computed from  the gluing of a  3-pt amplitude for a massive particle of  spin $s$ minimally coupled to gravity, with the graviton helicities to be  $\pm 2$, and a massless 3-pt amplitude two massless legs  of helicity  $(h,\mp h)$ and a  graviton leg of helicity $\mp 2$. Schematically we have 

\begin{figure}[h!]
\begin{equation}\label{eq:cuts}
  \includegraphics[width=78mm]{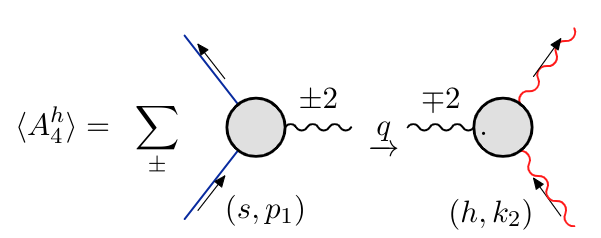}
\end{equation}
\end{figure}
\iffalse
\begin{equation}
 \begin{fmffile}{cutm4}
 \langle A_4^h\rangle=\hspace{0.2cm}\sum_{\pm}
 \parbox{62pt}
 {
  \parbox{15pt}{\begin{fmfgraph*}(59,65)
    \fmfstraight
    \fmfleft{i1,i2}
    \fmfright{o1}
    \fmf{plain,width=0.7,label= $(s,,p_1)$,foreground=(0.035,,0.168,,0.623)}{i1,v}
    \fmf{plain,width=0.7,foreground=(0.035,,0.168,,0.623)}{v,i2}
    \fmfv{decor.shape=circle,decor.filled=12,decor.size=.35w}{v}
    \fmf{photon,width=0.7,label= $\pm 2$,label.side=left,tension=1.5}{v,o1}
  \marrow{ea}{down}{top}{}{i1,v}
  \marrow{eo}{up}{top}{}{v,i2}
  \end{fmfgraph*}}}
\stackrel{}{\underrightarrow{q}}
\parbox{15pt}{
  \begin{fmfgraph*}(59,70)
    \fmfstraight
    \fmfright{i1,i2}
    \fmfleft{o1}
    \fmf{photon,width=0.7,label=$(h,,k_2)$,foreground=(1,,0.1,,0.1)}{v,i1}
    \fmf{photon,width=0.7,foreground=(1,,0.1,,0.1)}{v,i2}
    \fmfv{decor.shape=circle,decor.filled=12,decor.size=.35w}{v}
    \fmf{photon,width=0.7,label=$\mp 2$,label.side=right,tension=1.5}{v,o1}
    \marrow{ea}{down}{top}{}{i1,v}
    \marrow{eb}{up}{top}{}{v,i2}
  \end{fmfgraph*}}\hspace{0.3cm}\, .
\end{fmffile}
 \label{eq:cuts}
\end{equation}
\fi
%
The three-point amplitudes on the left hand side can be easily evaluated using spinor-helicity variables, see \cite{Guevara:2018wpp} for the details. Setting $q^\mu\sigma_\mu=|q]\langle q|$ for the on-shell transfer momentum we have
\begin{equation}
    \langle A_3^{s,+2}\rangle =\frac{\kappa}{2} \frac{\bra{r}p_1|q]^2}{\braket{qr}^2}  \,e^{- q{\cdot}a},\,\,\, \,\, \langle A_3^{s,-2}\rangle = \frac{\kappa}{2}\frac{\bra{q}p_1|r]^2}{[rq]^2}  \,e^{q{\cdot}a},
\end{equation}
with $\{\ket{r},  |r]\}$  some reference spinors.  On the other hand, amplitudes on the right hand side are totally fixed by little group covariance. Setting $k_2^\mu\sigma_\mu=|2]\langle 2|,\,\,k_3^\mu\sigma_\mu=|3]\langle 3|$ we have

\begin{equation}\label{eq:3ptmassls}
    A_3^{h,-h,+2} = \frac{\kappa}{2}\frac{[3q]^{2+2h}[q3]^{2-2h}}{[23]^2}, \,\, \,\,\, A_3^{h,-h,-2} = \frac{\kappa}{2}\frac{\braket{2q}^{2-2h}\braket{q3}^{2+2h}}{\braket{23}^2},   
\end{equation}
for the helicity preserving case. Analogous amplitudes can be written for the helicity reversing scenario. 
The t-channel residues are straightforward to compute following the prescription of \cref{sec:eikonal-scalars}. For the helicity conserving case, in the c.o.m. frame, we arrive at the amplitude 
\begin{equation}\label{eq:eikonal-amplitude-hconserving}
\begin{split}
    \langle A_4^{h,-h}\rangle &=-i \left(\frac{\kappa}{2}\right)^2\frac{(s-M^2)^{2-2h}}{t}\bra{3}p_1|2]^{2h}e^{q{\cdot}a},\\
  & = i(-iN^X)^{2h}\left(\frac{\kappa}{2}\right)^2 \frac{(2M\omega)^2}{\vec{q}^2}e^{i\vec{q}{\cdot}\hat{p}\times\vec{a}}, 
\end{split}    
\end{equation}
where
\begin{equation}
    -iN^X:=\frac{\langle 3|p_1|2]^{2h}}{(s-M^2)^{2h}}
\end{equation}
is a phase for real kinematics. As we mentioned in \cref{sec:eikonal-scalars}, this $h$-dependent phase can be remove by fixing the little group transformation and set to  $-iN^X=1$. Note that we recover the  amplitude \eqref{eq:eikonal-scalar}. 

In the QFT computation  we notice that  the $h$ dependent phase factor appearing in  the tree level amplitude  can be removed via a little group transformation.  This universality for the scattering angle was observed up to 2PM in the light-like scattering off a spinless background \cite{Bjerrum-Bohr:2016hpa}, where the scattering angle  was computed via  the cross section, which is independent of the phase factor.

Let us now comment on the  helicity reversing scenario. It is easy to show that the amplitude in this case  is given by 
\begin{equation}\label{eq:eq:eikonal-amplitude-hviolating}
\begin{split}
    \langle A_4^{h,h}\rangle &=-i \left(\frac{\kappa}{2}\right)^2\frac{(s-M^2)^{2-2h}}{t} M^{2h} [32]^{2h}e^{q{\cdot}a},\\
    \langle A_4^{h,h} \rangle  & =i\left(\frac{\kappa}{2}\right)^2 \frac{(2M\omega)^2}{\vec{q}^2}\left(\frac{[32]}{2\omega}\right)^{2h}\left[e^{i\vec{q}{\cdot}\hat{p}\times\vec{a}}+e^{-i\vec{q}{\cdot}\hat{p}\times\vec{a}}\right], 
\end{split}    
\end{equation}
and thus it is suppressed in the eikonal limit due to the extra factor of $[23]\sim t$, see also \cite{Brandhuber:2019qpg}.   

The suppression of the helicity reversing amplitude shows  that the $2\times 2$ scattering matrix defined in \eqref{eq:scattering-matrix-helicities} is proportional to the  identity, and hence there is no spin induced polarization in the eikonal limit.  This therefore  confirms that the non zero induced polarization computed in  \cref{sec:Section3} is indeed subleading in the eikonal limit. In addition, in \cite{Chen:2021huj} it  was observed that in eikonal limit for the scattering of two KBHs,   the spin-entanglement suppression can be trace back to  the suppression of spin-flipping components in the S-matrix. Here, \eqref{eq:eq:eikonal-amplitude-hviolating}  corresponds to the massless analog of the spin-flitting suppression, where indeed the helicity of the wave is conserved after scattering. 

The helicity independence of the scattering amplitude demonstrates the universality of the scattering angle and time delay computed in \cref{sec:eikonal-scalars}. One can check that taking the residue of the full amplitudes \eqref{scalar-kerr}, \eqref{neutrino-kerr}  in the eikonal limit, recovers amplitudes \eqref{eq:eikonal-amplitude-hconserving} and \eqref{eq:eq:eikonal-amplitude-hviolating}.

\bibliographystyle{JHEP}
\bibliography{references}

\end{document}